\documentclass[acmtog]{acmart}
\pdfoutput=1
\usepackage{booktabs} 

\usepackage{xr} 

\usepackage{geometrycollective}
\usepackage{xparse}
\usepackage{xspace}
\usepackage{multirow}

\usepackage{dsfont}

\usepackage{cancel}

\usepackage[ruled]{algorithm2e} 

\SetAlFnt{\small}
\SetAlCapFnt{\small}
\SetAlCapNameFnt{\small}
\SetAlCapHSkip{0pt}
\IncMargin{-\parindent}

\acmJournal{TOG}
\acmVolume{X}
\acmNumber{X}
\acmArticle{XX}
\acmYear{XXXX}
\acmMonth{-1}

\setcopyright{acmcopyright}

\acmDOI{XX}

\acmSubmissionID{XXX}

\begin{teaserfigure}
  \centering
   \includegraphics{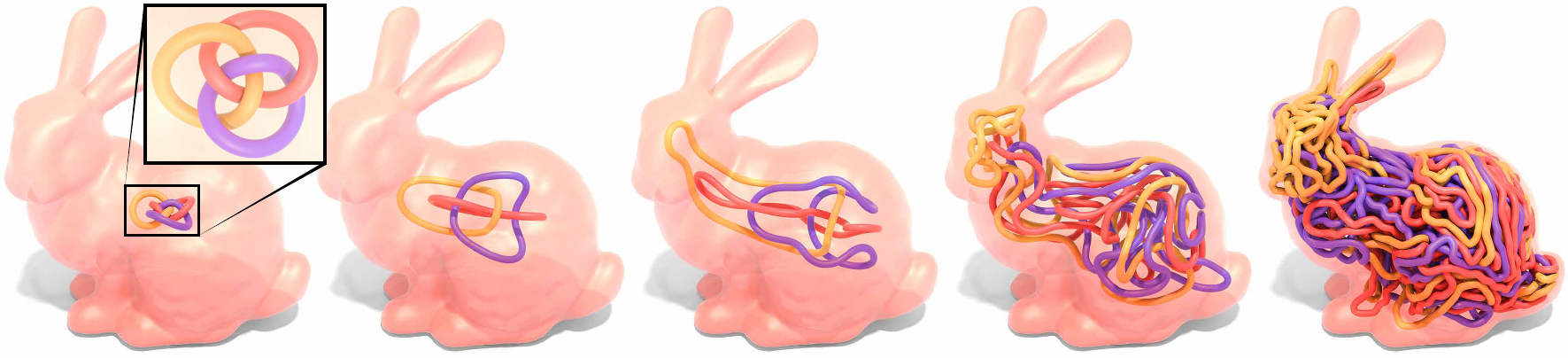}
   \caption{We develop an efficient strategy for optimizing curves while avoiding self-collisions.  Here for instance, interwoven curves of increasing length are confined inside a fixed domain, resulting in an intricate ``curve packing.''  Replacing ordinary gradient descent with a specially-tailored \emph{fractional Sobolev gradient} lets us take very large steps toward the solution, enabling rapid design exploration.\label{fig:BunnyTeaser}}
\end{teaserfigure}

\newcommand{\llangle}{\langle\!\langle} 
\newcommand{\rrangle}{\rangle\!\rangle} 

\newcommand{\Domain}{M} 

\newcommand{\RootNode}{\mathcal{R}}
\newcommand{\SourceNode}{\mathcal{B}}
\newcommand{\TargetNode}{\mathcal{A}}
\newcommand{\SourceEdge}{J}
\newcommand{\TargetEdge}{I}

\newcommand{\TargetIndex}{i}
\newcommand{\Center}{x}
\newcommand{\Tangent}{T}
\newcommand{\Mass}{\ell}
\newcommand{\Node}{\mathcal{N}}
\newcommand{\NodeCenter}{\overline{x}}
\newcommand{\NodeTangent}{\overline{T}}
\newcommand{\NodeTangentPoint}{\overline{p}}
\newcommand{\NodeMass}{L}

\newcommand{\Index}[2]{#1[#2]}

\newcommand{\ones}{\mathbf{1}}

\newcommand{\low}{0}

\newcommand{\shortminus}{\text{--}}

\newcommand{\DerivativeMatrix}{\mathsf{D}} 
\newcommand{\SobolevMatrix}{\mathsf{A}} 
\newcommand{\VSobolevMatrix}{\overline{\mathsf{A}}} 
\newcommand{\ConstraintMatrix}{\mathsf{C}} 
\newcommand{\HighOrderMatrix}{\mathsf{B}} 
\newcommand{\LowOrderMatrix}{\mathsf{B}^{\low}} 

\newcommand{\SSinner}{{H^s_\gamma}} 

\providecommand{\T}{\top}

\DeclareDocumentCommand{\Grass}{ O{\DomDim} O{\AmbDim} }{
	\operatorname{Gr}_{#1}^{#2}
}

\newcommand{\Ecal}{\mathcal{E}}
\newcommand{\hEcal}{\hat{\Ecal}}

\newcommand{\DirichletEnergy}{\Ecal_D}

\DeclareDocumentCommand{\DiscEnergy}{ O{\NumExp} O{\DenomExp} }{
   \hEcal^{#1}_{#2}
}

\DeclareDocumentCommand{\Energy}{ O{\NumExp} O{\DenomExp} }{
   \smash{\Ecal^{#1}_{#2}}
}

\DeclareDocumentCommand{\Kernel}{ O{\NumExp} O{\DenomExp} }{
	k^{#1}_{#2}
}
\DeclareDocumentCommand{\DiscKernel}{ O{\NumExp} O{\DenomExp} }{
        \hat{k}^{#1}_{#2}
}
\DeclareDocumentCommand{\AmbKernel}{ O{\NumExp} O{\DenomExp} }{
	\varPhi^{#1}_{#2}
}

\DeclareDocumentCommand{\Sobo}{ O{} O{} o O{\R}}{
	\IfValueTF{#3}{
	  W^{#1}_{#2}(#3;#4)
	}{
	  W^{#1}_{#2}
	}
}
\DeclareDocumentCommand{\SoboC}{ O{} O{} }{\Sobo[#1][#2][\Circle][\AmbSpace]}

\DeclareDocumentCommand{\Holder}{ O{} O{} o O{\R}}{
	\IfValueTF{#3}{
	  C^{#1}_{#2}(#3;#4)
	}{
	  C^{#1}_{#2}
	}
}
\DeclareDocumentCommand{\HolderC}{ O{} O{} }{\Holder[#1][#2][\Circle][\AmbSpace]}

\DeclareDocumentCommand{\Lebesgue}{ O{} O{} o O{\R}}{
	\IfValueTF{#3}{
	  L^{#1}_{#2}(#3;#4)
	}{
	  L^{#1}_{#2}
	}
}
\DeclareDocumentCommand{\LebesgueC}{ O{} O{} }{\Lebesgue[#1][#2][\Circle][\AmbSpace]}

\newcommand{\NumExp}{\alpha}
\newcommand{\DenomExp}{\beta}

\newcommand{\id}{\operatorname{id}}

\newcommand{\DomDim}{n}
\newcommand{\AmbSpace}{\R^\AmbDim}
\newcommand{\AmbDim}{3}

\newcommand{\diag}{\operatorname{diag}}

\newcommand{\R}{\mathbb{R}}

\newcommand{\ceq}{\coloneqq}

\newcommand{\cD}{\mathcal{D}}

\usepackage{braket}
\usepackage[draft]{fixme}

\begin{document}
\title{Repulsive Curves}
\author{Chris Yu}
\affiliation{%
  \institution{Carnegie Mellon University}
  }
\author{Henrik Schumacher}
\affiliation{%
  \institution{RWTH Aachen University}
  \streetaddress{Templergraben 55}
  \city{Aachen, Germany}
  \postcode{52062}
  }
\author{Keenan Crane}
\affiliation{%
  \institution{Carnegie Mellon University}
  \streetaddress{5000 Forbes Ave}
  \city{Pittsburgh}
  \state{PA}
  \postcode{15213}}

\renewcommand\shortauthors{Yu, Schumacher, and Crane}

\begin{abstract}
   Curves play a fundamental role across computer graphics, physical simulation, and mathematical visualization, yet most tools for curve design do nothing to prevent crossings or self-intersections.  This paper develops efficient algorithms for (self-)repulsion of plane and space curves that are well-suited to problems in computational design.  Our starting point is the so-called \emph{tangent-point energy}, which provides an infinite barrier to self-intersection.  In contrast to local collision detection strategies used in, \eg, physical simulation, this energy considers interactions between all pairs of points, and is hence useful for global shape optimization: local minima tend to be aesthetically pleasing, physically valid, and nicely distributed in space.  A reformulation of gradient descent, based on a \emph{Sobolev-Slobodeckij inner product} enables us to make rapid progress toward local minima---independent of curve resolution.  We also develop a hierarchical multigrid scheme that significantly reduces the per-step cost of optimization.  The energy is easily integrated with a variety of constraints and penalties (\eg, inextensibility, or obstacle avoidance), which we use for applications including curve packing, knot untangling, graph embedding, non-crossing spline interpolation, flow visualization, and robotic path planning.
\end{abstract}

%
%


%
%


\thanks{%
}


\maketitle

\section{Introduction}
\label{sec:Introduction}

Shape optimization plays a role in a broad range of tasks ranging from variational data fitting to computational design.  However, for many tasks it is essential to design \emph{in context}, \ie, relative to the geometry of the surrounding environment.  Hard boundary conditions (\eg, fixing the endpoints of a cable) provide a basic mechanism for providing context, but do not account for another fundamental requirement: physical objects cannot penetrate solid objects in the environment, nor can they intersect themselves.  In some contexts, self-intersection can be avoided by detecting and resolving collisions at the moment of impact.  However, forward simulation is not particularly effective at guiding shape optimization toward an intelligent design---for example, untangling a complicated knot via forward physical simulation is just as hard as trying to untangle it by hand.  In this paper we instead explore how a global variational approach to curve self-avoidance provides new opportunities for computational design.

\setlength{\columnsep}{1em}
\setlength{\intextsep}{0em}
\begin{wrapfigure}[7]{r}{72pt}
   \includegraphics{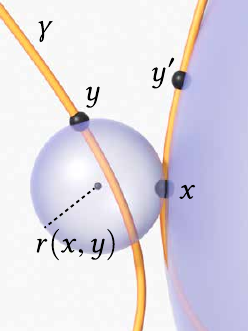}
\end{wrapfigure}
Our starting point is the \emph{tangent-point energy} of \citet{Buck:1995:ASE}, which for an arc-length parameterized curve \(\gamma: M \to \RR^\AmbDim\) can be expressed as an integral over all pairs of points \((x,y) \in M^2 := M \times M\):

\vspace{.5\baselineskip}
\begin{minipage}{.6\columnwidth}
\begin{equation}
   \label{eq:TangentPointEnergy}
   \boxed{\mathcal{E} := \iint_{\Domain^2} \frac{1}{r(\gamma(x),\gamma(y))^\alpha}\ dx dy.}
   \vspace{.5\baselineskip}
\end{equation}
\end{minipage}

\noindent Here \(r(x,y)\) is the radius of the smallest sphere tangent to \(\gamma(x)\) and passing through \(\gamma(y)\), and \(\alpha \in \RR\) is a parameter controlling the strength of repulsion.  This energy approaches infinity for points \(\gamma(y)\) that are close to \(\gamma(x)\) in space but far from \(\gamma(x)\) along the curve itself---preventing self-collision.  For points \(\gamma(y^\prime)\) close to \(\gamma(x)\) \emph{along} the curve, the radius \(r\) is very large---keeping forces bounded, and making the integral well-defined.

\newpage

Although this energy has a simple definition, its gradient involves high-order \emph{fractional} derivatives.  Hence, classic optimization techniques must take extremely small steps, and standard techniques from shape optimization are not well-suited to handle the nonlocal nature of the energy.  Our approach is to develop a preconditioner that \emph{exactly} matches the fractional order of the differential (\secref{Optimization}).  In doing so, we obtain a gradient descent equation involving no spatial derivatives, permitting large time steps that make rapid progress toward local minima (\figref{FreedmanComparison}).  In practice, this method is orders of magnitude more efficient than the simple untangling schemes often used in the knot literature (\figref{OtherMethodsComparison}), and offers substantial improvements over general-purpose optimization techniques from geometry processing (\secref{EvaluationAndComparisons}).  Algorithms of this flavor have proven effective for problems such as finding minimal surfaces~\cite{Pinkall:1993:CDM}, integrating Willmore flow~\cite{Schumacher:2017:HGF}, and computing surface parameterizations~\cite{Kovalsky:2016:AQP}.  However, little work has been done in the more challenging setting of nonlocal, ``all-pairs'' energies.

\paragraph{Contributions.}

Though knot energies have received significant attention in mathematics, there has been relatively little work on numerical tools for computational design.  In this paper we develop:
\begin{itemize}
   \item a principled discretization of the tangent-point energy,
   \item a novel preconditioner based on the \emph{Sobolev-Slobodeckij} inner product,
   \item a numerical solver that easily incorporates constraints needed for design, and
   \item a Barnes-Hut strategy and hierarchical multigrid scheme for the tangent-point energy that greatly improve scalability.
\end{itemize}
We also explore a collection of constraints and potentials that enable us to apply this machinery to a broad range of applications in visualization and computational design (\secref{ResultsAndApplications}).

\section{Related Work}
\label{sec:RelatedWork}

We briefly review topics related to computational design of curves; \secref{CurveEnergies} gives more detailed background on curve energies.  At a high level, computational design of free-form curves has generally focused on specific domains such as road networks~\cite{Hassan:1998:SAT,McCrae:2009:SPC}, telescoping structures~\cite{Yu:2017:CDT}, or rod assemblies~\cite{Perez:2015:DFF,Zehnder:2016:DSS}; \citet[Chapter 3]{Moreton:1992:MCV} gives a history of traditional design via spline curves.  Our goal is to develop tools that can be applied to a wide range of multi-objective design scenarios, as explored in \secref{ResultsAndApplications}.

\subsection{Curve Simulation}
\label{sec:CurveSimulation}

One natural idea is to avoid collision via physics-based simulation of elastic rods~\cite{Bergou:2008:DER}.  However, the paradigm of collision detection and response is ``too local'': for computational design, one aims to globally optimize a variety of design criteria, rather than simulate the behavior of a given curve.  \emph{Sensitivity analysis}, which provides sophisticated \emph{local} improvement of an initial design, has been successfully applied to several rod design problems~\cite{Perez:2015:DFF,Zehnder:2016:DSS,Perez:2017:CDA}.  This technique can be seen as complementary to global repulsion-based form-finding, helping to incorporate, \eg{}, nonlinear mechanical phenomena into a final design.  Curves also arise naturally as filaments or field lines in continuum phenomena like fluids, plasmas, and superfluids~\cite{Angelidis:2005:SSB,Weissmann:2010:FBS,Padilla:2019:BRI,Kleckner:2016:HSV,Chern:2016:SS,DeForest:2007:FML}.  However, using such phenomena for curve design is challenging since (i) initial conditions are hard to construct, and (ii) these systems naturally exhibit \emph{reconnection events} where distinct pieces of a curve merge~\cite{Maucher:2016:UKR}.

\subsection{Knot Energies}
\label{sec:KnotEnergies}

\begin{figure}[t]
   \centering
   \includegraphics{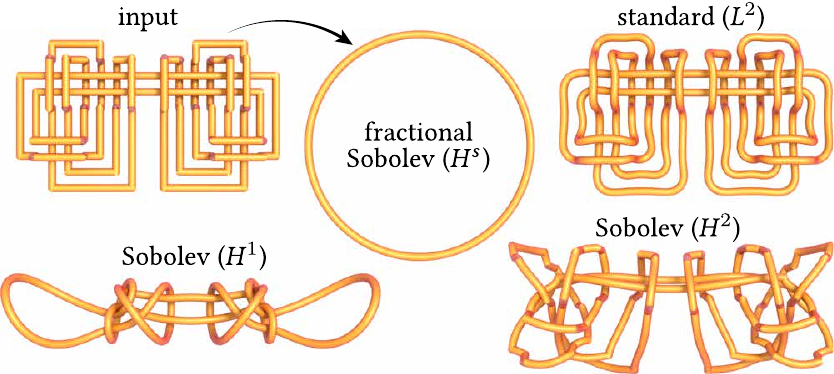}
   \caption{Untangling the \emph{Freedman unknot} \figloc{(top left)} to the unit circle.  For the same wall clock time, standard \(L^2\) gradient descent makes almost no progress, whereas conventional Sobolev descent fails to smooth out low \((H^1)\) or high \((H^2)\) frequencies.  By carefully matching the inner product to the energy, our fractional \(H^s\) descent quickly flows to the circle.\label{fig:FreedmanComparison}}
\end{figure}

Motivated by questions in mathematics, biology, and physics~\cite{Calvo:2002:PK}, there is a significant body of work on the \emph{unknot problem}: can a closed loop be continuously deformed into a circle without passing through itself (\ie, via \emph{isotopy})?  Solving this decision problem is not our goal---so far it is not clear it can even be done in polynomial time~\cite{Lackenby:2014:ENT}.  Yet knot untangling energies (discussed in \secref{CurveEnergies}) provide a valuable starting point for computational design.  Numerically, simple ad-hoc methods that repel all pairs of vertices can yield inconsistent, unreliable behavior and slow convergence (\figref{OtherMethodsComparison}, \figloc{right}).  Starting with more principled discretizations, \emph{KnotPlot}~\cite{Scharein:1998:ITD} uses a simple relaxation scheme, and \citet{Kusner:1998:Mobius} apply a standard conjugate gradient method via \emph{SurfaceEvolver}~\cite{Brakke:1992:SE}, both evaluating all \(O(n^2)\) interactions between the \(n\) vertices.  Other, adjacent methods have been developed for \emph{tightening} a given knot~\cite{Pieranski:1998:SIK,Ashton:2011:KTC}, simulating the knot tying process~\cite{Brown:2004:RTK,Kubiak:2007:RMR,Harmon:2009:ACM}, or untangling knots without optimizing their shape~\cite{Ladd:2004:MPK}; more recent methods apply \(L^2\)~\cite{Walker:2016:SOS} or integer Sobolev (\(H^2\)) descent~\cite{Bartels:2018:SSA}.  Octrees have been used to evaluate the ropelength of a static knot~\cite{Ashton:2005:FOB}, but Barnes-Hut/multipole schemes have not yet been developed for energy minimization.  Likewise, little has been said about fractional preconditioners, and treatment of general constraints.

Our approach builds on careful analysis of the fractional Sobolev spaces associated with the tangent point energy~\cite{Blatt:2012:BRE,Blatt:2013:EST,Blatt:2015:RTT}. Whereas this work focuses on, \eg{}, the existence of local minimizers and short-time existence of gradient flows in the smooth setting, we use it to develop numerical algorithms.

\subsection{Geometric Optimization}
\label{sec:GeometricOptimization}

Optimization of curve and surface energies can be greatly accelerated by ``Sobolev-like'' preconditioning.  The idea is to replace the ordinary \(L^2\) inner product with one that is carefully matched to the energy, yielding a gradient flow that is much easier to integrate (\secref{WarmUpDirichletEnergy} gives a didactic example).  Such flows make more rapid progress toward minimizers (\figref{FreedmanComparison}), since energy is reduced uniformly across all spatial frequencies.  Importantly, Sobolev preconditioners are most effective when the \emph{order of the preconditioner is perfectly matched to the order of spatial derivatives in the energy}.  A preconditioner whose order is too high or too low can slow down convergence---see for instance \figref{SobolevDirichlet}, \figloc{bottom-right}.

Sobolev-type preconditioners have seen some prior use in geometry processing.  For example, the minimal surface algorithm of \citet{Pinkall:1993:CDM} effectively performs Sobolev descent \cite[Section 16.10]{Brakke:1994:SEM}, but was not originally framed in these terms; \citet{Renka:1995:MSS} give an algorithm directly formulated via a (variable) Sobolev inner product.  Later work adopts Sobolev-like strategies for surface fairing and filtering~\cite{Desbrun:1999:IFI,Eckstein:2007:GSF,Martin:2013:ENO,Crane:2013:RFC,Schumacher:2017:HGF}.  More recently Sobolev-like descent has become a popular strategy for minimizing elastic energies, such as those that arise in surface parameterization and shape deformation~\cite{Kovalsky:2016:AQP,Claici:2017:IAP,Zhu:2018:BCQ}.  See \secref{EvaluationAndComparisons} for more in-depth discussion and comparisons.

\vspace{\baselineskip}

Importantly, previous work does not consider the challenging \emph{fractional} case, which differs significantly from standard Sobolev preconditioning.  From an analytical point of view, one must do work even to determine the order of derivatives arising in the differential, which we do by reasoning about the associated function spaces (\appref{SobolevSlobodeckijGradient}).  We use this knowledge to formulate a novel preconditioner in the smooth setting which carefully considers lower-order terms (\secref{Optimization}), which we then translate into the discrete setting via a principled discretization of the tangent-point energy (\secref{Discretization}).  From a computational point of view, the machinery needed to apply a fractional preconditioner is also different from ordinary Sobolev preconditioners: one cannot simply solve a sparse linear system, but must instead construct an efficient hierarchical scheme for (approximately) inverting a dense nonlocal operator.  None of these pieces appear in the previous work discussed above.  Moreover, existing Sobolev preconditioners (such as those based on the Laplacian) and standard optimization strategies (such as Newton descent) are not as effective for our problem---as we show via extensive numerical experiments (\secref{EvaluationAndComparisons}).

\section{Curve Energies}
\label{sec:CurveEnergies}

We first give a detailed discussion of the tangent-point energy, which we optimize in \secref{Optimization}.  Throughout we will use single bars \(|X|\) and brackets \(\langle X, Y \rangle\) to denote the Euclidean inner product on vectors in \(\RR^\AmbDim\), and reserve double bars \(\|f\|\) and brackets \(\llangle f, g \rrangle\) for norms and inner products on functions.  We also use \(\cdot |_f\) to indicate that a quantity (\eg{}, an energy) is evaluated at a function \(f\).

\subsection{Background}
\label{sec:Background}

Consider a collection of curves given by a parameterization \(\gamma: \Domain \to \RR^3\), where \(M\) is comprised of intervals and/or loops.  How can we formulate an energy that prevents self-intersection of \(\gamma\)?  In general we will consider energies of the form
\[
   \mathcal{E}(\gamma) = \iint_{\Domain^2} k(x,y)\ dx_\gamma dy_\gamma,
\]
where the kernel \(k: \Domain \times \Domain \to \RR\) captures the interaction between two points on the curve, and \(dx_\gamma\) denotes the length element on \(\gamma\).

\subsubsection{Electrostatic Potential}
\label{sec:ElectrostaticPotential}

One natural idea for defining \(k\) is to imagine that there is electric charge distributed along \(\gamma\) that pushes it away from itself, producing the Coulomb-like potential
\begin{equation}
   \label{eq:CoulombPotential}
   k_{\text{Coulomb}}(x,y) := \frac{1}{|\gamma(x)\!-\!\gamma(y)|^\alpha},
\end{equation}
\setlength{\columnsep}{1em}
\setlength{\intextsep}{0em}
\begin{wrapfigure}{r}{48pt}
   \vspace{-\baselineskip}
   \includegraphics[width=48pt]{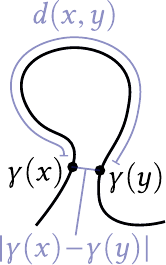}
\end{wrapfigure}
where the parameter \(\alpha\) controls the strength of repulsion.  Unfortunately this simple energy does not work for a continuous curve: for \(\alpha < 2\) it is not strong enough to prevent collisions, allowing the curve to pass through itself---yet for \(\alpha \geq 1\) the integral does not exist, resulting in unpredictable and unreliable behavior when discretized.

\subsubsection{M\"{o}bius Energy}
\label{sec:MobiusEnergy}

\begin{figure}[b]
   \includegraphics{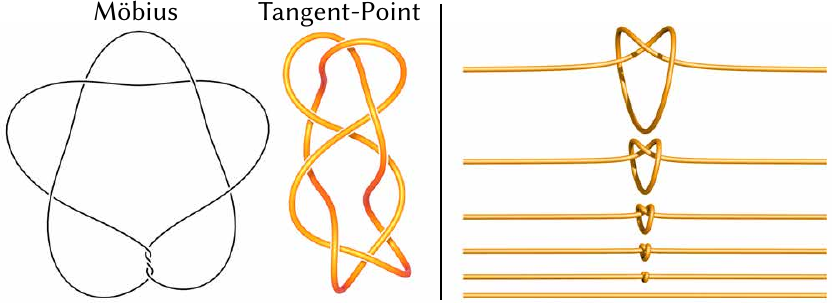}
   \caption{\figloc{Left:} Since the M\"{o}bius energy is scale-invariant, it allows ``tight spots'' where the curve nearly touches itself; such features are avoided by the tangent-point energy.  \figloc{Right:} The M\"{o}bius energy can likewise artificially eliminate knots by pulling them tight at no energetic cost. (Leftmost image from \citet{Kusner:1998:Mobius}.)\label{fig:TightSpot}}
\end{figure}

To obtain a well-defined energy, one can \emph{regularize} the integrand in regions where \(x\) approaches \(y\).  One such regularization, proposed by \citet{OHara:1991:EAK}, is the \emph{M\"{o}bius energy}, with kernel
\[
   k_{\text{M\"{o}bius}}(x,y) := \frac{1}{|\gamma(x)\!-\!\gamma(y)|^2} - \frac{1}{d(x,y)^2},
\]
where \(d(x,y)\) denotes the shortest distance between \(x\) and \(y\) \emph{along} the curve (\eg, the smaller of two arcs along a circle).  Intuitively: if two points are both close in space and close along the curve, we remove the singular energy; if they are close in space but \emph{distant} along the curve, they continue to repel each other (see inset).  This energy is invariant to M\"{o}bius transformations~\cite{Freedman:1994:MEK}, which can be attractive from the perspective of knot theory---but causes problems for computational design, since near-intersections may not be penalized in a natural way (\figref{TightSpot}).

\subsection{Tangent Point Energy}
\label{sec:TangentPointEnergy}

Instead, we will use the tangent point energy introduced in \secref{Introduction}.  We can write this energy more explicitly by noting that
\[
   r(x,y) = \frac{|\gamma(x)\!-\!\gamma(y)|^2}{|T(x) \times (\gamma(x)\!-\!\gamma(y))|}
\]
where \(T(x)\) is the unit tangent of \(\gamma\) at \(x\). This expression leads to a generalized tangent-point energy~\cite{Blatt:2015:RTT}, given by
\[
   \Energy(\gamma) := \iint_{\Domain^2} \Kernel (\gamma(x),\gamma(y),T(x) )\ dx_\gamma dy_\gamma,
\]
where \(\Kernel\) is the \emph{tangent-point kernel}
\begin{equation}
   \label{eq:TangentPointKernel}
   \Kernel(p,q,T) := \frac{|T \times (p-q)|^\NumExp}{|p-q|^\DenomExp}.
\end{equation}
In the case \(\beta = 2\alpha\), this energy agrees with \eqref{TangentPointEnergy}; as shown by \citet{Blatt:2013:EST} it is well-defined for any \(\alpha,\beta\) satisfying \(\alpha > 1\) and \(\beta \in [\alpha+2,2\alpha+1)\) (\lemref{FiniteTangentPointEnergy}).  Most importantly, it tends toward infinity as the curve approaches itself, preventing self-intersection.  In particular, when \(\beta-\alpha > 2\) it is not scale-invariant, and hence avoids the pull-tight phenomenon. (We set \((\alpha,\beta)\) to \((2,4.5)\) in Figures \ref{fig:Dataset}--\ref{fig:ScatterPlots}, and \((3,6)\) elsewhere.)

\begin{figure}
   \includegraphics{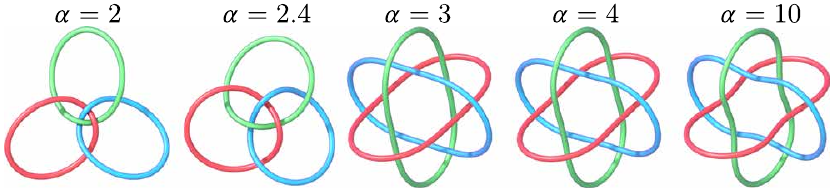}
   \caption{Local minimizers of the tangent-point energy \(\Energy[\alpha][2\alpha]\).  When \(\alpha=2\) the tangent-point energy is scale-invariant and can exhibit ``tight spots''; for larger values of \(\alpha\) local interactions are penalized more than distant ones.\label{fig:BorromeanRings}}
\end{figure}

This energy is also attractive for design since it provides natural regularization, akin to bending energy.  The reason is that the integrand can vanish only for a straight line (where the radius \(r\) is infinite at every point).  The powers \(\DenomExp\) and \(\NumExp\) have an impact on this bending behavior---for instance, if \(\DenomExp = 2\NumExp\), then a higher \(\NumExp\) gives a more repulsive energy where curves are willing to bend more in order to avoid collision (\figref{BorromeanRings}).

\section{Optimization}
\label{sec:Optimization}

Consider an energy \(\Ecal\) that depends on a function \(f\).  A typical starting point for optimization is to integrate the gradient flow
\begin{equation}
   \label{eq:GradientFlow}
   \tfrac{d}{dt} f = -\grad \Ecal(f),
\end{equation}
\ie, to move in the direction of ``steepest descent.'' As mentioned in \secref{RelatedWork}, however, the efficiency of this flow depends critically on the \emph{inner product} used to define the gradient---in other words, there are many different notions of what it means to be ``steepest.'' Recall in particular that the \emph{differential} \(d\Ecal\) describes the change in \(\Ecal\) due to any small perturbation \(u\) of \(f\):
\[
   d\Ecal|_{f}(u) = \lim_{\varepsilon \to 0} \tfrac{1}{\varepsilon}\left( \Ecal(f + \varepsilon u) - \Ecal(f) \right).
\]
The \emph{gradient} of \(\Ecal\) is then the unique function \(\grad\Ecal\) whose inner product with any function \(u\) gives the differential in that direction:
\begin{equation}
   \label{eq:Gradient}
   \llangle \grad \Ecal, u \rrangle_V = d\Ecal(u).
\end{equation}
Traditionally, the inner product \(\langle\!\langle \cdot, \cdot \rangle\!\rangle_V\) is just the \(L^2\) inner product
\[
   \llangle u, v \rrangle_{L^2} := \textstyle \int_M \langle u(x), v(x) \rangle\ dx.
\]
More generally, however, one can try to pick a so-called \emph{Sobolev inner product} \(\llangle u, v \rrangle_{H^k}\) that yields an easier gradient flow equation.  Examples include the \(H^1\) and \(H^2\) inner products, which for a domain without boundary can be written as
\begin{equation}
   \label{eq:H1InnerProduct}
   \llangle u, v \rrangle_{H^1} := \llangle \grad u, \grad v \rrangle_{L^2} = -\llangle \Delta u, v \rrangle_{L^2},
\end{equation}
and
\begin{equation}
   \label{eq:H2InnerProduct}
   \llangle u, v \rrangle_{H^2} := \llangle \Delta u, \Delta v \rrangle_{L^2} = \llangle \Delta^2 u, v \rrangle_{L^2},
\end{equation}
which measure first and second derivatives (\resp{}) rather than function values.  In general, if we write our inner product as \(\llangle u, v \rrangle_{H^k} = \llangle Au, v \rrangle_{L^2}\) for some linear operator \(A\), then we can express the new gradient direction \(g\) as the solution to
\begin{equation}
   \label{eq:GradientTransformation}
   Ag = \grad_{L^2}\mathcal{E}.
\end{equation}
This transformation is akin to the preconditioning provided by Newton's method, except that we replace the Hessian with an operator \(A\) that is always positive-definite, and often easier to invert.  In particular, when \(A\) comes from a carefully-designed Sobolev inner product, it will eliminate spatial derivatives, avoiding the stringent time step restriction typically associated with numerical integration of gradient flow (\figref{FlowTrajectories}).

\subsection{Warm-up: Dirichlet energy}
\label{sec:WarmUpDirichletEnergy}

\begin{figure}[b]
   \centering
   \includegraphics{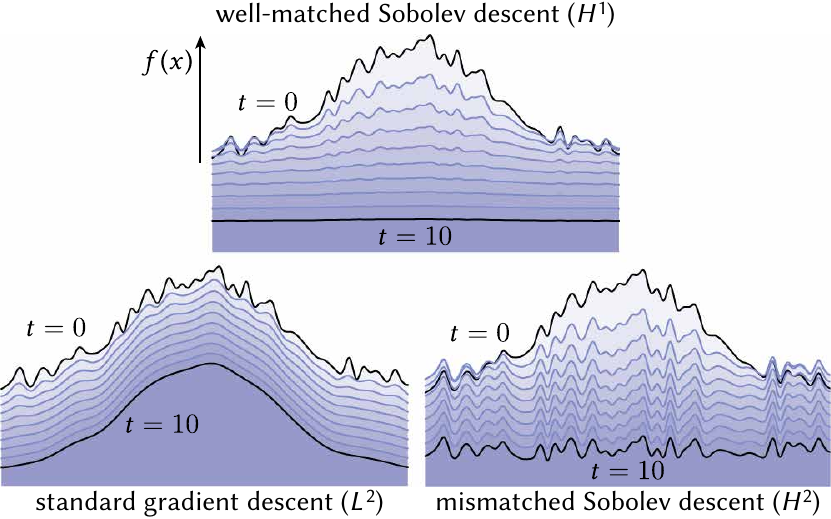}
   \caption{For Dirichlet energy, which penalizes variations in a function \(f(x)\), standard \(L^2\) gradient descent mostly smooths out local features \figloc{(bottom left)}, whereas an inner product that is too high-order has trouble removing high frequencies \figloc{(bottom right)}.  A Sobolev descent that is well-matched to the order of the energy yields rapid progress toward a local minimizer \figloc{(top)}.  We apply a similar strategy to quickly optimize the shape of curves.\label{fig:SobolevDirichlet}}
\end{figure}

Since analysis of the tangent-point energy is quite involved, we begin with a standard ``toy'' example that helps sketch out the main ideas of our approach.  In particular, consider the \emph{Dirichlet energy}
\begin{equation}
   \label{eq:DirichletEnergy}
   \DirichletEnergy(f) := \tfrac{1}{2} \textstyle \int_\Omega |\grad f(x)|^2\ dx,
\end{equation}
which penalizes variation in a function \(f: \Omega \to \RR\).  If the domain \(\Omega\) has no boundary, then we can use integration by parts to write this energy as
\[
   \DirichletEnergy(f) = \tfrac{1}{2}\llangle \grad f, \grad f \rrangle_{L^2} = -\tfrac{1}{2}\llangle \Delta f, f \rrangle_{L^2},
\]
where \(\Delta\) denotes the Laplace operator.  The differential is then
\[
   d\DirichletEnergy|_f(u) = -\llangle \Delta f, u \rrangle_{L^2},
\]
and from \eqref{Gradient}, we see that the \(L^2\) gradient of \(\DirichletEnergy\) is given by \(\grad_{L^2} \DirichletEnergy|_f = -\Delta f\). Hence, \(L^2\) gradient descent yields the \emph{heat flow}
\[
   \tag{\(L^2\) gradient flow}
   \tfrac{d}{dt} f = \Delta f,
\]
which involves second-order derivatives in space~\cite[Section 1.2]{Andrews:2020:EGF}.  If we try to solve this equation using, say, explicit finite differences with grid spacing \(h\), we will need a time step of size \(O(h^2)\) to remain stable---significantly slowing down computation as the grid is refined.  To lift this time step restriction, we can use a different inner product to define the gradient.  In particular, replacing \(\langle\!\langle \cdot, \cdot \rangle\!\rangle_V\) with the \(H^1\) inner product in \eqref{Gradient} yields
\begin{equation}
   \label{eq:DirichletH1Gradient}
   \llangle \Delta \grad_{H^1}\! \DirichletEnergy, u \rrangle_{L^2} = \llangle \Delta f, u \rrangle_{L^2}.
\end{equation}
This equation can be satisfied by letting \(\grad_{H^1}\! \DirichletEnergy := f\), in which case \eqref{GradientFlow} defines an \(H^1\) gradient flow
\[
   \tag{\(H^1\) gradient flow}
   \tfrac{d}{dt} f = -f.
\]
\setlength{\columnsep}{1em}
\setlength{\intextsep}{0em}
\begin{wrapfigure}{r}{120pt}
   \includegraphics{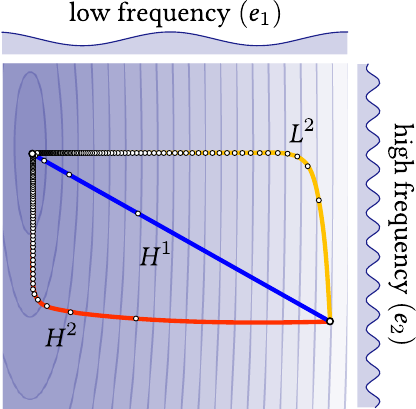}\vspace{-.5\baselineskip}
   \caption{Gradient flows projected onto a low- and high-frequency mode \(e_1, e_2\), \resp{}  Notice that poor preconditioning leads to slow convergence.\label{fig:FlowTrajectories}}
\end{wrapfigure}
This flow involves no spatial derivatives, and hence comes with no time step restriction.  In effect, rather than a PDE, we now have a system of independent ODEs, which is far easier to integrate numerically.  As shown in \figref{SobolevDirichlet}, the character of this flow is quite different: it makes progress by simultaneously flattening all spatial frequencies, rather than just performing local smoothing.  While this approach is not appropriate for dynamical simulation, it is quite useful for finding local minima, as needed in geometric design.  In general, however, Sobolev descent is not as simple as just uniform scaling---instead, one must solve a linear PDE (\eqref{GradientTransformation}) for the new descent direction.

Note that we should not use an inner product with \emph{too many} derivatives.  For example, if we use the \(H^2\) inner product (\eqref{H2InnerProduct}) we get a gradient \(\grad_{H^2} \DirichletEnergy|_f = -\Delta^{-1} f\), and a flow
\[
   \tag{\(H^2\) gradient flow}
   \tfrac{d}{dt} f = \Delta^{-1} f.
\]
This flow is again hard to integrate, and has trouble smoothing out high frequencies (\figref{SobolevDirichlet}, \figloc{bottom-right}).  In general, one cannot achieve good behavior by blindly picking a Sobolev inner product, but must instead \emph{carefully match the inner product to the energy}.

\paragraph{Low-Order Terms}  One remaining issue is that \eqref{DirichletH1Gradient} determines the \(H^1\) gradient only up to functions in the null space of \(\Delta\).  This situation is problematic, since it means we cannot obtain a gradient by solving \eqref{GradientTransformation} directly (with \(A = -\Delta\)).  Instead, we must include \emph{low-order terms} that make the overall operator \(A\) invertible.  For instance, we could let \(A := -\Delta + \id\), where \(\id\) denotes the identity.  But if we uniformly scale the domain by a factor \(c > 0\), the new operator looks like \(-\tfrac{1}{c^2}\Delta + \id\) and the character of the flow changes substantially: when \(c\) is small it looks like the \(H^1\) flow; when \(c\) is large, it looks more like the \(L^2\) flow.  Careful treatment of regularization and scaling is therefore an important consideration in the development of our curve flow (\secref{LowOrderTerm}).

\subsection{Fractional Sobolev Gradient}
\label{sec:FractionalSobolevGradient}

In the case of a nonlocal energy like the tangent-point energy \(\Energy\), one can no longer use a standard Sobolev inner product---instead, an inner product of \emph{fractional} order is needed, in order to match fractional derivatives that appear in the differential.  Construction of a suitable inner product for the tangent-point energy is fairly technical---in a nutshell, we begin with a known expression for the \emph{fractional Laplacian} on Euclidean \(\RR^\DomDim\), and formulate an analogous operator for embedded curves.  Taking additional (integer) derivatives yields a differential operator \(B_\sigma\) of the same order as the differential \(d\Energy\).  We then add a lower-order operator \(B^\low_\sigma\) that makes the overall operator \(A_\sigma := B_\sigma + B^{\low}_\sigma\) more well-behaved.  Our \emph{Sobolev-Slobodeckij inner product} is then defined as
\[
   \llangle u, v \rrangle_{\SSinner} := \llangle A_\sigma u, v \rrangle_{L^2}.
\]
Details are given in \appref{SobolevSlobodeckijGradient}---here we give only the most essential definitions needed to derive our discrete algorithm (\secref{Discretization}).

\subsubsection{Derivative Operator}
\label{sec:DerivativeOperator}

To define the inner product, we will need the first derivative operator \(\cD\) given by
\begin{equation}
   \label{eq:DerivativeOperator}
   \cD u := du\, d\gamma^\T / |d\gamma|^2.
\end{equation}
This operator just takes the usual derivative of \(u\) along \(\Domain\) and expresses it as a vector in \(\RR^3\) tangent to \(\gamma\); the factor \(1/|d\gamma|^2\) accounts for the fact that the curve is not in general arc-length parameterized.

\subsubsection{High-Order Term}
\label{sec:HighOrderTerm}

As discussed in \appref{OrderOfTheDifferential}, the differential \(d\Energy\) of the tangent-point energy has order \(2s\), where \(s = (\beta-1)/\alpha\).  To build an inner product of the same order, we first define the fractional differential operator \(B_\sigma\), given by
\begin{equation}
   \label{eq:HighOrderTerm}
   \llangle B_\sigma u, v \rrangle := \!\!\iint_{\Domain^2} \!\!
   	\frac{\cD u(x)\!-\!\cD u(y)}{|\gamma(x)\!-\!\gamma(y)|^\sigma} \frac{\cD v(x)\!-\!\cD v(y)}{|\gamma(x)\!-\!\gamma(y)|^\sigma}  \frac{dx_\gamma dy_\gamma}{|\gamma(x)\!-\!\gamma(y)|}
\end{equation}
for all sufficiently regular \(u,v: \Domain \to \RR\), where \(\sigma = s - 1\).
This operator also has order \(2s\) (\appref{FractionalInnerProduct}), and plays a role analogous to the Laplacian in \secref{WarmUpDirichletEnergy}.  Yet just like the Laplacian, \(B_\sigma\) is only \emph{semidefinite}, since it vanishes for functions that are constant over each component of the domain \(\Domain\).  Hence, it is not invertible, and cannot be used directly to solve for a descent direction---instead we must ``regularize'' \(B_\sigma\) by adding an additional, lower-order term.

\subsubsection{Low-Order Term}
\label{sec:LowOrderTerm}

\begin{figure}[t!]
   \includegraphics[width=\columnwidth]{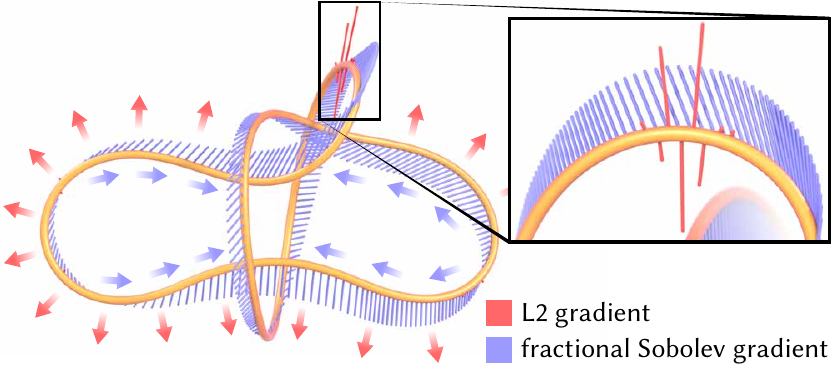}
   \caption{Since an \(L^2\) gradient flow is always perpendicular to the curve \figloc{(red)}, it fails to resolve even simple cases like the one shown above, where a large near-tangential motion is needed to untangle a knot.  The fractional Sobolev gradient \figloc{(blue)} permits such motions, yielding a far more efficient flow.\label{fig:L2HardCase}}
\end{figure}

A na\"{i}ve approach to regularization, like adding some small \(\epsilon > 0\) times the identity, yields undesirable behavior---\(\varepsilon\) must be sufficiently large to have an effect, but if \(\varepsilon\) is too large, motion is significantly damped.  Moreover, an inner product constructed this way will no longer exhibit predictable scaling behavior, \ie, rescaling the input will actually change the \emph{direction} of the gradient rather than just its magnitude---and hence can change the solution obtained by a designer.  Instead, we carefully choose an additional, low-order term \(B^{\low}_\sigma\) that not only provides the right scaling behavior, but also enables us to steer the flow more quickly toward self-avoiding configurations (\figref{L2HardCase}).  In particular, we add the term \(\llangle B^{\low}_\sigma u, v \rrangle\), given by
\begin{equation}
   \label{eq:LowOrderTerm}
   \iint_{\Domain^2} \Kernel[2][4] (\gamma(x),\gamma(y),T(x)) \frac{(u(x)\!-\!u(y))(v(x)\!-\!v(y))}{|\gamma(x)\!-\!\gamma(y)|^{2\sigma+1}} \ dx_\gamma dy_\gamma,
\end{equation}
where \(\Kernel\) is the tangent-point kernel given in \eqref{TangentPointKernel}.  See \appref{FractionalInnerProduct} for further discussion.

\subsubsection{Sobolev-Slobodeckij Gradient}
\label{sec:Sobolev-SlobodeckijGradient}

Following \eqref{Gradient}, our final gradient \(\grad_\SSinner\) is defined via the fractional inner product:
\begin{equation}
   \label{eq:SSGradient}
   \llangle \grad_\SSinner \Energy, X \rrangle_\SSinner = d\Energy|_{\gamma}(X), \quad \text{for all}\ X: \Domain \to \RR^3.
\end{equation}
Since \(\grad_\SSinner \Energy\) and \(X\) are vector- rather than scalar-valued, we apply the inner product componentwise.  In other words,
 \begin{equation}
    \label{eq:SoboSloboGradient}
    \grad_{\SSinner}\Energy = \bar{A}_\sigma^{-1} \grad_{L^2}\Energy|_{\gamma},
 \end{equation}
where \(\bar{A}_\sigma\) denotes componentwise application of \(A_\sigma\).  Note that the combined operator \(A_\sigma = B_\sigma + B^\low_\sigma\) still has \emph{globally} constant functions in its kernel, corresponding to global translations. To make \eqref{SoboSloboGradient} well-defined, we can simply add any constraint that fixes the translation of the curve (\secref{Constraints}).  In practice, we never need a closed-form expression for the gradient, nor do we explicitly invert the operator \(A_\sigma\); instead, we solve \eqref{GradientTransformation} numerically.

\section{Discretization}
\label{sec:Discretization}

We now use the inner product from the previous section to derive an efficient numerical scheme for minimizing the tangent-point energy.  The description given here assumes a na\"{i}ve implementation using dense matrices and an \(O(n^2)\) evaluation of the energy and its differential; hierarchical acceleration is described in \secref{Acceleration}.

\begin{figure}[t]
   \includegraphics{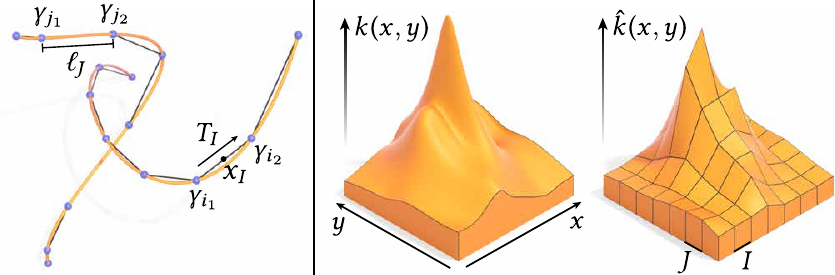}
   \caption{\figloc{Left:} notation used for discrete curves. \figloc{Right:} Our discrete energy is obtained by applying the trapezoidal rule to the smooth energy for each edge pair \(I,J\).\label{fig:Discretization}}
\end{figure}

\paragraph{Notation} In the discrete setting, we will model any collection of curves and loops (including several curves meeting at a common point) as a graph \(G = (V,E)\) with vertex coordinates \(\gamma: V \to \RR^3\) (\figref{Discretization}); we use \(|V|\) and \(|E|\) to denote the number of vertices and edges, \resp{}\  For each edge \(I \in E\) with endpoints \(i_1,i_2\), we use
\[
   \ell_I := |\gamma_{i_1}-\gamma_{i_2}|, \quad
   \Tangent_I := (\gamma_{i_2}-\gamma_{i_1})/\ell_I, \quad \text{and} \quad
   \Center_I := (\gamma_{i_1}+\gamma_{i_2})/2
\]
to denote the edge length, unit tangent, and midpoint, \resp{}\  For any quantity \(u: V \to \RR\) on vertices we use \(u_I := (u_{i_1}+u_{i_2})/2\) to denote the average value on edge \(I = (i_1,i_2)\), and \(\usf[I] := [u_{i_1}\ u_{i_2}]^\T\) to denote the \(2 \times 1\) column vector storing the values at its endpoints.  Finally, we refer to any pair \((T,x) \in \RR^6\) as a \emph{tangent-point}.

\subsection{Discrete Energy}
\label{sec:DiscreteEnergy}

\setlength{\columnsep}{1em}
\setlength{\intextsep}{0em}
\begin{wrapfigure}{r}{71pt}
   \includegraphics[width=71pt]{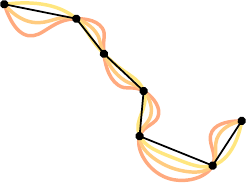}
\end{wrapfigure}
Since the tangent-point energy is infinite for polygonal curves \cite[Figure 2.2]{Strzelecki:2017:GCE}, we assume that \(\gamma\) is inscribed in some (unknown) smooth curve, and apply numerical quadrature to the smooth energy \(\Energy\).  The resulting discrete energy then approximates the energy of any sufficiently smooth curve passing through the vertices \(\gamma_i\).  We start by integrating \(\Kernel\) over all pairs of edges:
\begin{equation}
   \label{eq:PolygonalTangentPointEnergy}
   \sum_{I \in E} \sum_{J \in E} \int_{\bar{I}} \int_{\bar{J}} \Kernel(\gamma(x),\gamma(y),T_I)\ dx_\gamma dy_\gamma.
\end{equation}
Here \(\bar{I}\) denotes the interval along edge \(I\).  As stated, this expression is ill-defined since any two edges with a common endpoint contribute infinite energy.  One idea is to replace any such term with one proportional to the curvature of the circle passing through the three distinct endpoints (in the spirit of \eqref{TangentPointEnergy}).  However, such terms would contribute nothing to the energy in the limit of regular refinement (\figref{EnergyDiagonal})---hence, we simply omit neighboring edge pairs. Applying the (2D) trapezoidal rule to \eqref{PolygonalTangentPointEnergy} then yields a discrete energy
\begin{equation}
   \label{eq:DiscreteEnergy}
   \DiscEnergy(\gamma) = \mathop{\sum\sum}_{I,J \in E, I \cap J = \emptyset} (\DiscKernel)_{IJ} \ell_I \ell_J,
\end{equation}
where \(\hat{k}\) is the discrete kernel
\begin{equation}
   \label{eq:DiscreteKernel}
   (\DiscKernel)_{IJ} := {\textstyle \frac{1}{4} \sum_{i \in I} \sum_{j \in J} \Kernel(\gamma_i,\gamma_j,T_I)}.
\end{equation}
The discrete differential is then simply the partial derivatives of this energy with respect to the coordinates of all the curve vertices:
\[
   d\DiscEnergy|_\gamma = \left[ \begin{array}{ccc} \partial\Energy/\partial\gamma_1 & \cdots & \partial\Energy/\partial\gamma_{|V|} \end{array} \right] \in \RR^{3|V|}.
\]
These derivatives can be evaluated via any standard technique (\eg, by hand, or using symbolic or automatic differentiation).

\begin{figure}
   \includegraphics{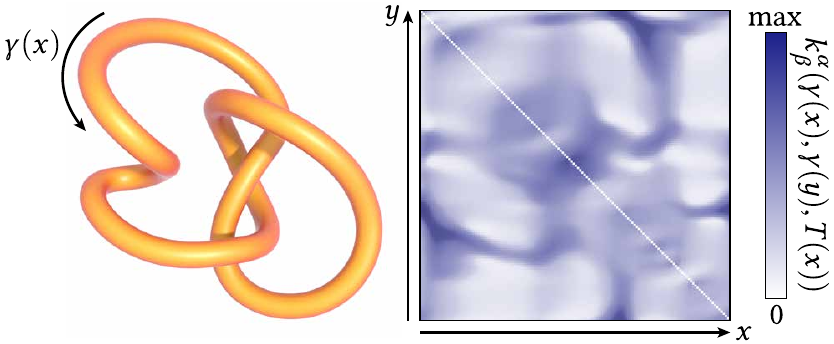}
   \caption{The tangent-point energy is a double integral of the kernel \(\Kernel\) \figloc{(right)} over the curve \(\gamma\) \figloc{(left)}.  Since this kernel is only weakly singular, omitting diagonal terms has an insignificant effect on the overall energy.
   \label{fig:EnergyDiagonal}}
\end{figure}

\subsection{Discrete Inner Product}
\label{sec:DiscreteInnerProduct}

As in the smooth setting, we define our inner product matrix as a sum \(\Asf = \Bsf + \Bsf^0\) of high-order and low-order terms \(\Bsf,\Bsf^0 \in \RR^{|V| \times |V|}\) (as defined below).  For \(\RR^3\)-valued functions, we also define a corresponding \(3|V| \times 3|V|\) matrix
\begin{equation}
   \label{eq:VectorInnerProduct}
   \VSobolevMatrix = \left[ \begin{array}{ccc} \SobolevMatrix & & \\ & \SobolevMatrix & \\ & & \SobolevMatrix \end{array} \right].
\end{equation}
Mirroring \eqref{GradientTransformation}, the discrete (fractional) Sobolev gradient \(\gsf \in \RR^{3|V|}\) is then defined as the solution to the matrix equation
\begin{equation}
   \label{eq:DiscreteGradientTransformation}
   \VSobolevMatrix \gsf = d\DiscEnergy.
\end{equation}

\subsubsection{Discrete Derivative Operator}
\label{sec:DiscreteDerivativeOperator}

For each edge \(I \in E\) we approximate the derivative \(\cD u\) of a function \(u: \Domain \to \RR\) (\eqref{DerivativeOperator}) via the finite difference formula \(\tfrac{1}{\ell_I}(\usf_{i_2}-\usf_{i_1}) T_I\), where \(\usf_i\) denotes the value of \(u\) sampled at vertex \(i\).  The corresponding derivative matrix \(\DerivativeMatrix \in \RR^{3|E| \times |V|}\) can be assembled from local \(3 \times 2\) matrices
\[
   \DerivativeMatrix_I = \tfrac{1}{\ell_I}[ \begin{array}{cc} -T_I & T_I \end{array} ].
\]

\subsubsection{Discrete High-Order Term}
\label{sec:DiscreteHighOrderTerm}

We approximate the high-order part of the inner product  \(\llangle B_\sigma u, v \rrangle\) as
\begin{equation}
   \label{eq:DiscreteHighOrderPart}
   \usf^\T \HighOrderMatrix \vsf = \mathop{\sum\sum}_{I,J \in E, I \cap J = \emptyset} w_{IJ} \langle \DerivativeMatrix_I \usf[I] - \DerivativeMatrix_J \usf[J], \DerivativeMatrix_I \vsf[I] - \DerivativeMatrix_J \vsf[J] \rangle,
\end{equation}
where the weights \(w_{IJ}\) arise from applying trapezoidal quadrature to the denominator in \eqref{FractionalPseudoLaplacian}:
\[
   w_{IJ} := \tfrac{1}{4} \ell_I \ell_J \textstyle\sum_{i \in I} \textstyle\sum_{j \in J} \frac{1}{|\gamma_i - \gamma_j|^{2\sigma+1}}.
\]
The entries of the corresponding Gram matrix \(\Bsf \in \RR^{|V| \times |V|}\) are obtained by differentiating \eqref{DiscreteHighOrderPart} with respect to the entries of \(\usf\) and \(\vsf\).  More explicitly, starting with the zero matrix one can build \(\Bsf\) by making the following increments for all pairs of disjoint edges \(I \cap J = \emptyset\), and all pairs of values \(a,b \in \{1,2\}\):
\[
   \arraycolsep=5pt
   \def\arraystretch{1.6}
   \begin{array}{ll}
      \HighOrderMatrix_{i_a i_b} +\!\!=\! (\shortminus{}1)^{a+b} w_{IJ} / \ell_I^2, &
      \HighOrderMatrix_{i_a j_b} -\!\!=\! (\shortminus{}1)^{a+b} w_{IJ} \,\langle T_I, T_J \rangle/(\ell_I \ell_J), \\
      \HighOrderMatrix_{j_a j_b} +\!\!=\! (\shortminus{}1)^{a+b} w_{IJ} / \ell_J^2, &
      \HighOrderMatrix_{j_a i_b} -\!\!=\! (\shortminus{}1)^{a+b} w_{IJ} \,\langle T_J, T_I \rangle/(\ell_J \ell_I).
   \end{array}
\]

\subsubsection{Discrete Low-Order Term}
\label{sec:DiscreteLowOrderTerm}

To discretize the low-order term \(B^{\low}_\sigma\) (\secref{LowOrderTerm}), we use a different discrete weight
\[
   w^{\low}_{IJ} := \tfrac{1}{4} \ell_I \ell_J \sum_{i \in I} \sum_{j \in J} \frac{\Kernel[2][4](\gamma_i,\gamma_j,T_I)}{|\gamma_i - \gamma_j|^{2\sigma+1}},
\]
and define a matrix \(\LowOrderMatrix \in \RR^{|V| \times |V|}\), given by the relationship
\[
   \usf^\T \LowOrderMatrix \vsf = \mathop{\sum\sum}_{I,J \in E, I \cap J = \emptyset} w^{\low}_{IJ} (\usf_I - \usf_J)(\vsf_I - \vsf_J).
\]
Following a similar derivation as above, this matrix can be constructed via the following increments:
\[
   \arraycolsep=10pt
   \def\arraystretch{1.6}
   \begin{array}{ll}
      \LowOrderMatrix_{i_a i_b} +\!= \tfrac{1}{4} w^{\low}_{IJ}, &
      \LowOrderMatrix_{i_a j_b} -\!= \tfrac{1}{4} w^{\low}_{IJ}, \\
      \LowOrderMatrix_{j_a i_b} -\!= \tfrac{1}{4} w^{\low}_{IJ}, &
      \LowOrderMatrix_{j_a j_b} +\!= \tfrac{1}{4} w^{\low}_{IJ}.
   \end{array}
\]

\subsection{Constraints}
\label{sec:Constraints}

\begin{figure}
   \centering
   \includegraphics{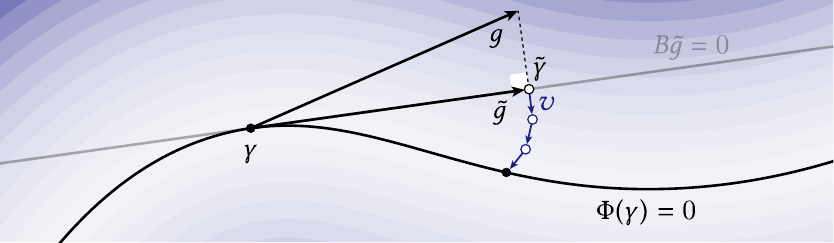}
   \caption{To enforce constraints \(\Phi(\gamma) = 0\) on the curve, we both project the gradient \(g\) onto the tangent of the constraint set, and also apply an iterative procedure to project the curve itself back onto the constraint set.  In both cases, the fractional Sobolev norm provides the definition of closeness.}
\end{figure}

For design applications, we will need to impose a variety of scalar constraints \(\Phi_i(\gamma) = 0\), \(i=1,\ldots,k\), which we encode as a single constraint function \(\Phi: \RR^{3 |V|} \to \RR^k\) (\secref{ConstraintsAndPotentials}).  To enforce these constraints, we project the gradient onto a valid descent direction (\secref{GradientProjection}); after taking a step in this direction, we also project the result onto the constraint set (\secref{ConstraintProjection}).

\subsubsection{Gradient Projection}
\label{sec:GradientProjection}

Let \(\ConstraintMatrix := d\Phi(\gamma)\) be the Jacobian matrix of the constraint, and let \(\gsf := \grad_{H^s_\gamma} E \in \RR^{3|V|}\) denote the unconstrained energy gradient.  We seek the descent direction \(\tilde{\gsf}\) that is closest to \(\gsf\) with respect to the fractional Sobolev norm, but which is also tangent to the constraint set:
\[
   \min_{\tilde{\gsf}} \tfrac{1}{2}|| \tilde{\gsf} - \gsf ||_{H^s_\gamma}^2 \quad \text{s.t.} \quad \ConstraintMatrix\tilde{\gsf} = 0.
\]
Writing \(||v||_{H^s_\gamma}^2\) as \(v^\T \VSobolevMatrix v\) (\secref{DiscreteInnerProduct}), we can apply the method of Lagrange multipliers to obtain the usual first-order optimality conditions, given by the saddle point system
\begin{equation}
\label{eq:GradientSaddlePointProblem}
   \left[
      \begin{array}{ll}
         \VSobolevMatrix & \ConstraintMatrix^\T \\
         \ConstraintMatrix & 0
      \end{array}
   \right]
   \left[
      \begin{array}{c}
         \tilde{\gsf} \\
         \uplambda
      \end{array}
   \right]
   =
   \left[
      \begin{array}{c}
         d\Energy|_\gamma^\T \\
         0
      \end{array}
   \right],
\end{equation}
where \(\uplambda \in \RR^k\) are the Lagrange multipliers, and we have applied the identity \(\VSobolevMatrix \gsf = d\Energy|_\gamma^\T\) (\eqref{DiscreteGradientTransformation}).

\subsubsection{Constraint Projection}
\label{sec:ConstraintProjection}

Suppose that we take a small step of size \(\tau\) along the projected gradient direction \(\tilde{\gsf}\) to get a new candidate curve \(\tilde{\gamma} := \gamma - \tau\tilde{\gsf}\).  To project this curve back onto the constraint set, we will apply an approximation of Newton's method that is faster to evaluate.  In particular, to find a displacement \(\xsf \in \RR^{3|V|}\) that takes us from \(\tilde{\gamma}\) back toward the constraint set \(\Phi(\gsf) = 0\), we solve the problem
\[
   \min_{\xsf} \tfrac{1}{2}\xsf^\T \VSobolevMatrix \xsf \quad \text{s.t.} \quad \ConstraintMatrix\xsf = -\Phi(\tilde{\gamma}).
\]
We then update our guess via \(\tilde{\gamma} \gets \tilde{\gamma} + \xsf\) and repeat until the constraint violation \(\Phi(\tilde{\gamma})\) is numerically small.  In practice, this process rarely takes more than three iterations.  At each iteration, \(\xsf\) is obtained by solving the saddle point problem
\begin{equation}
\label{eq:ConstraintSaddlePointProblem}
   \left[
      \begin{array}{ll}
         \VSobolevMatrix & \ConstraintMatrix^\T \\
         \ConstraintMatrix & 0
      \end{array}
   \right]
   \left[
      \begin{array}{c}
         \xsf \\
         \upmu
      \end{array}
   \right]
   =
   \left[
      \begin{array}{c}
         0 \\
         -\Phi(\tilde{\gamma})
      \end{array}
   \right],
\end{equation}
where \(\upmu \in \RR^k\) are Lagrange multipliers.

\subsection{Time Stepping}
\label{sec:TimeStepping}

A judicious choice of time step can significantly improve the efficiency of the flow.  One strategy is to use the first time step \(\tau_{\max}\) at which a collision occurs as the starting point for a line search, which guarantees that the curve remains in the same isotopy class.  (Similar approaches have been used in, \eg{}, \emph{KnotPlot}~\cite{Scharein:1998:ITD} for knot untangling, and by \citet{Smith:2015:BPF} for surface parameterization.) Computing this time step via standard techniques~\cite{Redon:2002:FCC} costs about as much as a single energy evaluation, \ie{}, significantly less than the overall cost of a single time step.  From here we apply standard backtracking line search \cite[Algorithm 9.2]{Boyd:2004:CO}; as a heuristic, we start this search at \(\tfrac{2}{3}\tau_{\max}\).  We use this strategy throughout \secref{EvaluationAndComparisons}.

An even simpler strategy that works well in practice (but comes with no collision guarantees) is to just normalize the gradient and perform backtracking line search starting with \(\tau=1\), until both (i) the Armijo condition is satisfied and (ii) constraint projection succeeds (\secref{ConstraintProjection}).  We use this latter strategy for all application examples in \secref{ResultsAndApplications}.  We stop when the \(L^2\) norm of the fractional Sobolev gradient goes below a user-specified tolerance \(\varepsilon\).  In our examples we use \(\varepsilon = 10^{-4}\), though of course for design applications one can also stop whenever the results are aesthetically pleasing.

\begin{figure}[t]
   \includegraphics[width=.8\columnwidth]{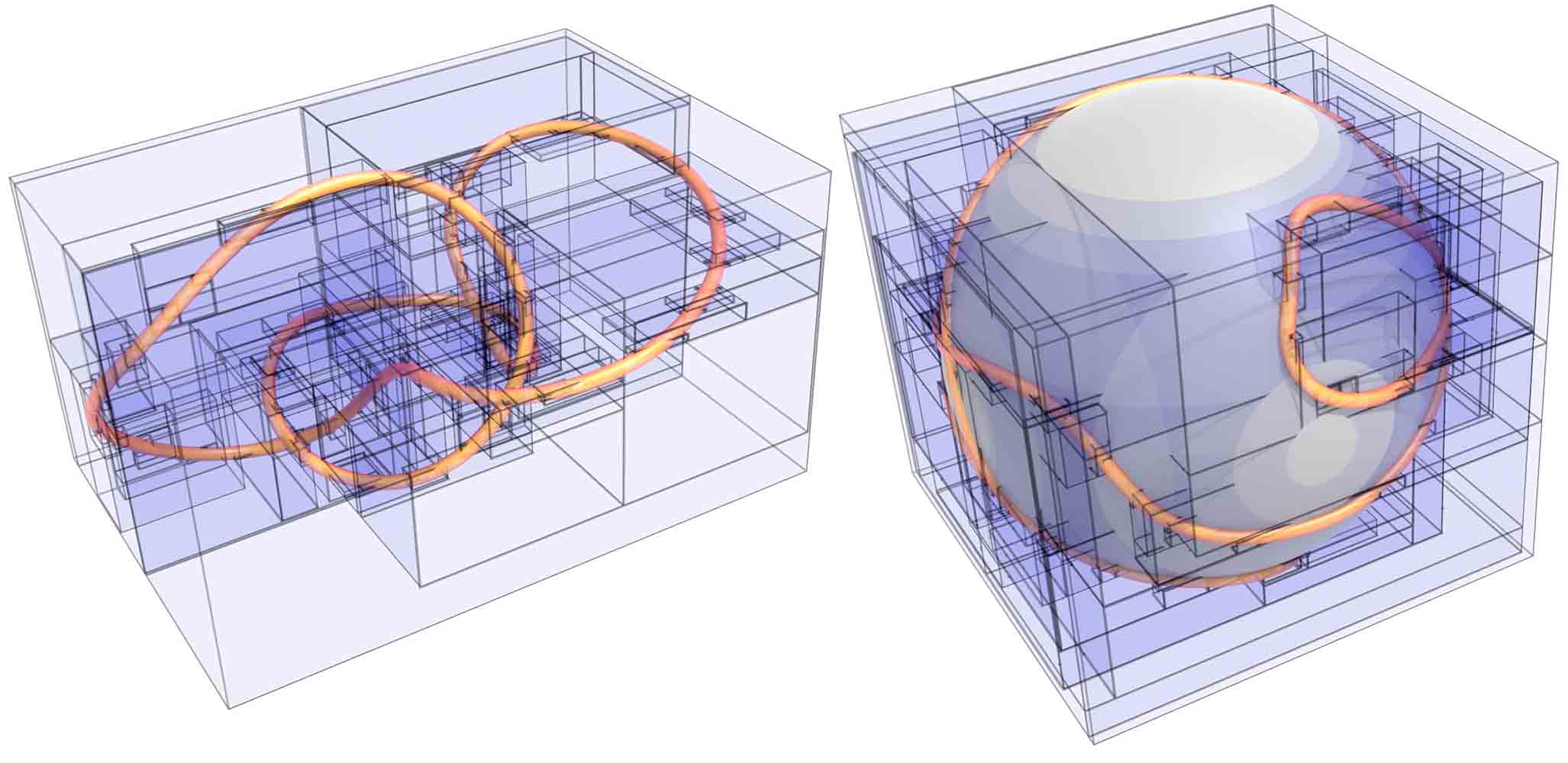}
   \caption{To accelerate evaluation of the tangent-point energy, we build a bounding volume hierarchy that partitions both positions \figloc{(left)} and tangent directions \figloc{(right)}, here drawn as a curve on the unit sphere.\label{fig:BoundingVolumeHierarchy}}
\end{figure}

\section{Acceleration}
\label{sec:Acceleration}

Computational design problems can entail large collections of curves with many thousands of vertices (\secref{ResultsAndApplications}).  Optimization hence becomes expensive since it involves not only an all-pairs energy (\secref{DiscreteEnergy}), but also inverting a dense inner product (\secref{DiscreteInnerProduct}).  However, since the kernel falls off rapidly in space, we can use hierarchical approximation to avoid a \(\Omega(|V|^2)\) time and storage cost.  Though our high-level approach is reasonably standard, careful consideration of the tangent-point energy is needed to develop a scheme that is efficient, easy to implement, and handles general nonlinear constraints.  To streamline exposition, we reserve the details of this scheme for \appref{AccelerationScheme}; at a high level it consists of three main parts, outlined below.  Note that since we care only about finding a good descent direction---and not accurately simulating a dynamical trajectory---we are free to use low-order schemes, which still provide good preconditioning.  Empirically, the overall strategy exhibits near-linear scaling in both time and memory (\figref{HierarchicalScaling}).

\subsection{Energy and Differential Evaluation}
\label{sec:EnergyandDifferentialEvaluation}

To accelerate evaluation of the energy \(\smash{\DiscEnergy}\) and its differential, we apply the \emph{Barnes-Hut algorithm} from \(N\)-body simulation \cite{BarnesHFCA1986}.  The basic idea is to approximate distant energy contributions by aggregating values in a spatial hierarchy.  In our case, this hierarchy must have six dimensions rather than three, since \(\smash{\DiscEnergy}\) depends on both positions \(\gamma \in \RR^3\) and tangents \(T \in \RR^3\).  In lieu of a standard octree we therefore use an axis-aligned \emph{bounding volume hierarchy (BVH)}, for which additional dimensions do not incur significant cost (\figref{BoundingVolumeHierarchy}).  \appref{EnergyAndDifferentialEvaluation} gives further details.

\subsection{Hierarchical Matrix-Vector Product}
\label{sec:HierarchicalMatrix-VectorProduct}

\begin{figure}[b]
   \vspace{-.75\baselineskip}\includegraphics{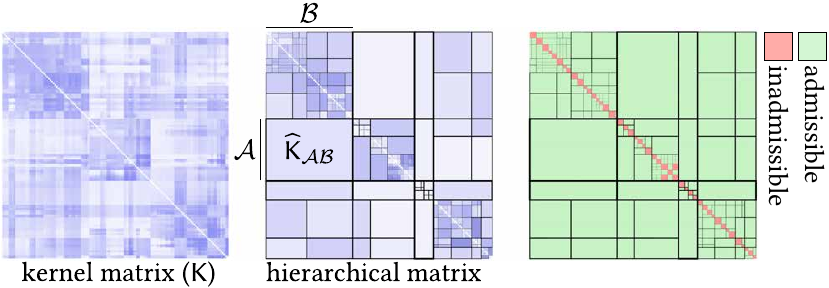}
   \vspace{-.25\baselineskip}\caption{\figloc{Left:} A kernel matrix \(\mathsf{K}\) encodes interactions between all pairs of edges. \figloc{Center:} To accelerate multiplication, this matrix is approximated by rank-1 blocks \(\widehat{K}_{\TargetNode\SourceNode}\), corresponding to pairs \((\TargetNode,\SourceNode)\) of distant BVH nodes.  \figloc{Right:} For pairs that are too close, this approximation is \emph{inadmissible}, and we must use the original matrix entries.\label{fig:BlockClusterTree}}
\end{figure}

For optimization we need to solve linear systems involving so-called \emph{kernel matrices}.  Any such matrix \(\Ksf \in \RR^{|E| \times |E|}\) has a special form
\[
   \Ksf_{\TargetEdge\SourceEdge} = k(p_\TargetEdge,p_\SourceEdge)\ell_\TargetEdge \ell_\SourceEdge,
\]
where the kernel \(k\) maps a pair of tangent-points to a real value (\secref{CurveEnergies}).  If \(k\) is a sufficiently regular, then \(\Ksf\) is well-approximated by a \emph{hierarchical matrix}~\cite{Hackbusch:2015:HMA}, \ie, a matrix of low-rank blocks (\figref{BlockClusterTree}).  Encoding this matrix as a \emph{block cluster tree (BCT)} enables fast matrix-vector multiplication via the \emph{fast multipole method}~\cite{Greengard:1997:NVF}.  Like the BVH, our BCT involves both positions \emph{and} tangents; in fact, each BCT block corresponds to a pair of BVH nodes.  See \appref{HierarchicalMatrixVectorProduct} for details.

\begin{figure}[t]
   \includegraphics{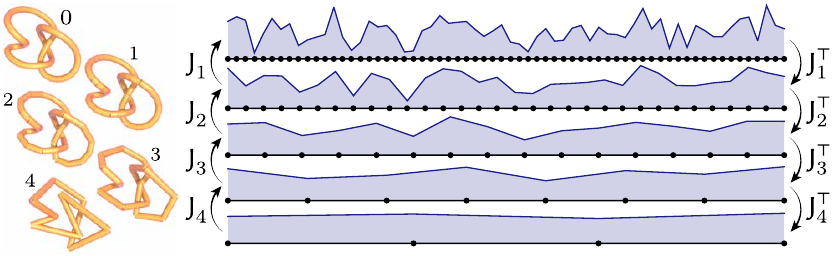}
   \caption{We accelerate linear solves using multigrid on a hierarchy of curves.\label{fig:Multigrid}}
\end{figure}

\subsection{Multigrid Solver}
\label{sec:MultigridSolver}

Since the hierarchical matrix-vector multiply does not build an explicit matrix, we use an iterative method to solve our linear systems. Empirically, off-the-shelf methods such as \(GMRES\) and \(BiCGStab\) are not well-suited for our problem. Instead, we use geometric multigrid (\figref{Multigrid}), since (i) it is straightforward to coarsen a curve network, and (ii) the low frequency modes of our Laplace-like operators are well-captured on a coarse mesh.  In the Euclidean case, this type of approach has been used successfully by \citet{Ainsworth:2017:AAF}.  \appref{MultigridSolver} describes our geometric coarsening/prolongation operators, as well as our multigrid strategy for both Sobolev gradient evaluation and constraint projection.

\section{Evaluation and Comparisons}
\label{sec:EvaluationAndComparisons}

We performed extensive evaluation and comparisons of our fractional Sobolev descent strategy relative to other methods.  Here we give an overview of results; a detailed account of how these evaluations were performed can be found in supplemental material.

\begin{figure}[b]
   \includegraphics{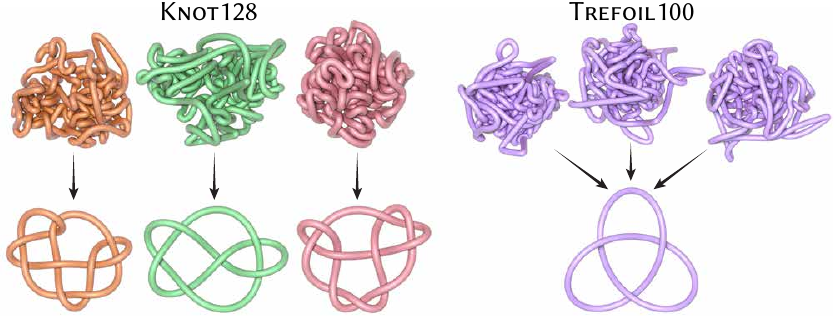}
   \caption{To evaluate performance, we built a ``stress test'' dataset of 128 random embeddings of different knot classes \figloc{(left)} and 100 random embeddings of the trefoil knot \figloc{(right)}.  The tangent point energy drives these curves toward much simpler embeddings, as shown here.\label{fig:Dataset}}
\end{figure}

\subsection{Dataset} We created two datasets of difficult knot embeddings: \textsc{Knot128}, which contains random embeddings of 128 distinct isotopy classes from \emph{KnotPlot}'s ``knot zoo,''  and \textsc{Trefoil100}, which contains 100 random embeddings of the trefoil knot (\figref{Dataset}).  We also used the \emph{Freedman unknot} (\figref{FreedmanComparison}, \figloc{top left}), which is a standard ``challenge problem'' from the knot energy literature~\cite[Section 3.3]{Scharein:1998:ITD}.  To examine scaling under refinement, we performed regular refinement on knots from each of these sets.

\subsection{Performance Comparisons} We compared our fractional Sobolev descent strategy to a variety of methods from optimization and geometry processing.  Overall, methods that use our fractional preconditioner performed best, especially as problem size increases.  We first ran all methods on several resolutions of a small set of test curves (\figref{Multiresolution}); we then took the fastest methods, and ran them on all 228 curves from our two datasets (\figref{ScatterPlots}).  For simplicity we did not use hierarchical acceleration in our method (and instead just solve dense systems), which gave a \emph{significant} performance advantage to alternative methods (which are based on sparse solves).  Even with this handicap, the fractional approach outperformed all other methods; as indicated in \figref{HierarchicalScaling}, hierarchical acceleration would widen this gap even further.  Importantly, previous methods also have a much higher failure rate at untangling difficult curves (\figref{ScatterPlots}).  Note that some previous methods do not directly handle hard nonlinear constraints; for these methods we perform an apples-to-apples comparison by replacing---in \emph{all} methods---hard edge length constraints with a soft elastic penalty (see supplemental material for further details).

\begin{figure}[b]
   \includegraphics{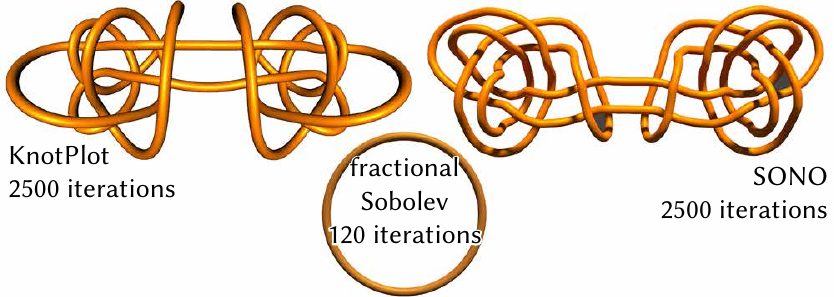}
   \caption{Our fractional Sobolev strategy is dramatically more efficient than previous methods for knot untangling---here we untangle the unknot from \figref{FreedmanComparison}. Neither KnotPlot nor SONO converged after several hours.\label{fig:OtherMethodsComparison}}
\end{figure}

\paragraph*{Knot untangling methods} We first compared to two well-known methods for knot untangling (\figref{OtherMethodsComparison}):  \emph{KnotPlot}, based on the so-called \emph{symmetric energy}, and \emph{shrink on no overlaps (SONO)}~\cite{Pieranski:1998:SIK} which performs a local iterative projection in the spirit of contemporary \emph{position-based dynamics}~\cite{Muller:2007:PBD}.  Both methods successfully untangle the Freedman knot, but only after tens of thousands of iterations~\cite[Figure 7.6]{Scharein:1998:ITD}. The basic reason is that, like \(L^2\) descent, such methods focus on reduction of local error, making global convergence quite slow.

\begin{figure}
   \includegraphics{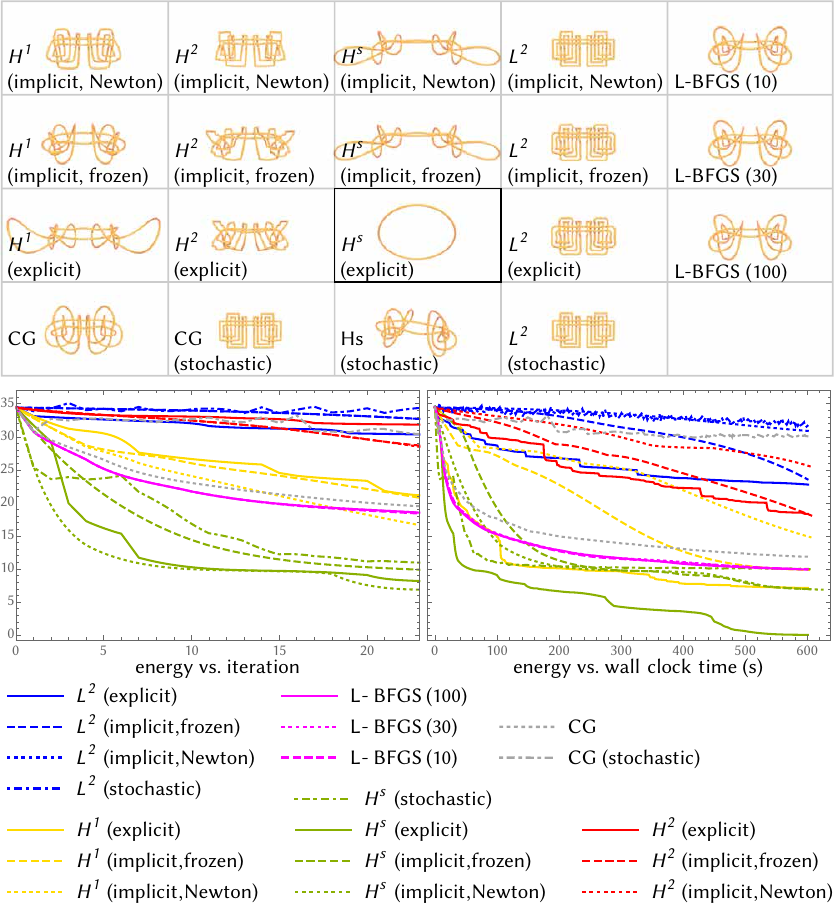}
   \caption{Across a wide variety of descent methods and inner products, our fractional Sobolev approach does significantly better both in terms of energy reduction per iteration \figloc{(middle left)} and real-world run time \figloc{(middle right)}.  At top we show results for an equal amount of compute time.\label{fig:PerformanceComparison}}\vspace{-\baselineskip}
\end{figure}

\paragraph*{1st-order methods} \figref{PerformanceComparison} indicates that basic 1st-order schemes like ordinary \(L^2\) gradient descent, L-BFGS using 10, 30, or 100 vectors, and nonlinear conjugate gradients \ala{}~\citet{Fletcher:1964:FMC} exhibit poor performance relative to our fractional scheme in terms of both wall clock time and number of iterations.  This example also indicates that for \(1 < s < 2\), the next smallest or largest \emph{integer} Sobolev preconditioners (\(H^1\) and \(H^2\)) underperform the fractional \(H^s\) preconditioner, whether using explicit or implicit Euler.  Since the implicit update equations are nonlinear, we solved them using Newton's method---either by updating the Hessian for each Newton step, or ``freezing'' the Hessian at the beginning of the time step.  Here we also tried stochastic gradient descent (SGD) with respect to the \(L^2\) inner product, which is currently quite popular in machine learning---this method did far worse than any other scheme we tried.  SGD with respect to \(H^s\) works better, but the speedup provided by stochastic evaluation does not compensate for the poor quality of the descent direction. 

\paragraph*{2nd-order methods} Second-order schemes like Newton's method can be adapted to nonconvex problems by projecting the Hessian onto a nearby positive-semidefinite matrix.  Since a global projection is prohibitively expensive, a heuristic sometimes used in geometric optimization is to project and sum up the Hessians of each local energy term~\cite{Teran:2005:RQF}; in our case we can decompose the energy into the edge-edge terms from \eqref{DiscreteEnergy}.  Though this heuristic can work well for, \eg{}, elastic energies, it does not appear to work very well for the tangent-point energy, and for larger examples had among the slowest run times of any scheme we tried (\figref{Multiresolution}).

\paragraph*{Quasi-Newton methods} Several recent methods from geometry processing apply Sobolev-like preconditioning to elastic energies, such as those used for shape deformation or surface parameterization~\cite{Kovalsky:2016:AQP,Claici:2017:IAP,Zhu:2018:BCQ}.  Since the highest-order term in such problems often looks like Dirichlet energy, \(H^1\) preconditioning via the Laplacian \(\Delta\) can be an effective starting point for optimization (as discussed in \secref{WarmUpDirichletEnergy}).  However, such preconditioners do not perform as well as our fractional preconditioner, since they are not as well-matched to the order of the differential \(d\Energy\). For instance, as seen in \figref{Multiresolution}, the AQP strategy of \citet{Kovalsky:2016:AQP} significantly underperforms our preconditioner when the Laplacian is used as the quadratic proxy; using our fractional operator as the quadratic proxy improves performance---but of course requires the machinery introduced in this paper. Another possibility is to use Laplacian-initialized L-BFGS (in the spirit of \citet{Zhu:2018:BCQ}); we found this strategy works a bit better than AQP, but again not as well as the fractional preconditioner.  We also considered several variants of these strategies, such as applying Nesterov acceleration, and combining nonlinear conjugate gradients (NCG) \ala{} \citet{PolakRibiere} or L-BFGS with our fractional preconditioner.  For hard constraints we advocate the use of our fractional (\(H^s\)) projected gradient scheme (as detailed in \secref{Discretization}); if soft constraint enforcement is acceptable, then L-BFGS or \(H^s\)-preconditioned NCG are both good options: the former converges faster near minima; the latter gets stuck less often.

\begin{figure}
   \includegraphics{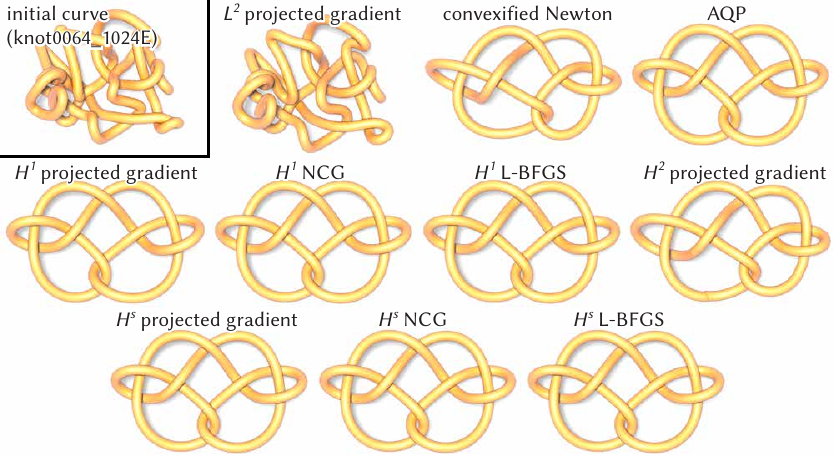}
   \caption{The tangent point energy appears to have relatively few local minima; hence, different descent strategies tend to find the same local minimizers (though some, like \(L^2\), do not find solutions in a reasonable amount of time).  See supplemental material for several hundred more examples.\label{fig:LocalMinima}}
\end{figure}

\begin{figure*}
   \includegraphics{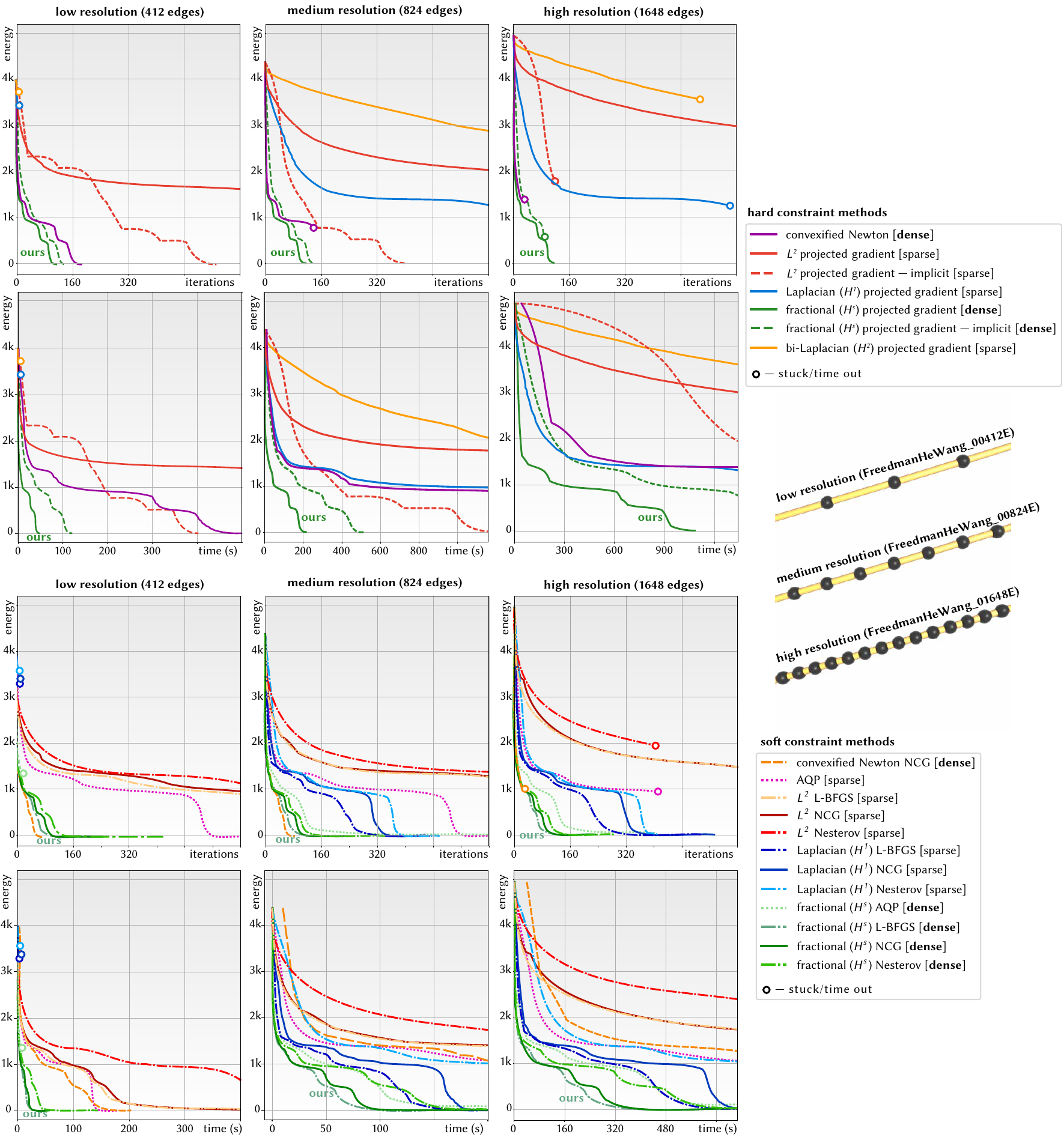}
   \caption{We compared our descent strategy to a variety of 1st-order, 2nd-order, and quasi-Newton strategies, using both hard constraints (top) and a soft penalty (bottom) to preserve length.  Here we show energy versus both time and iteration count for several resolutions of the initial curve from \figref{FreedmanComparison}; tests on additional curves yield very similar results (see supplemental material).  Note that we achieve the best real-world clock time---even though we compare a dense implementation of our method (without hierarchical acceleration) to sparse versions of other schemes.\label{fig:Multiresolution}}
\end{figure*}

\begin{figure*}
   \vspace{5\baselineskip}\includegraphics{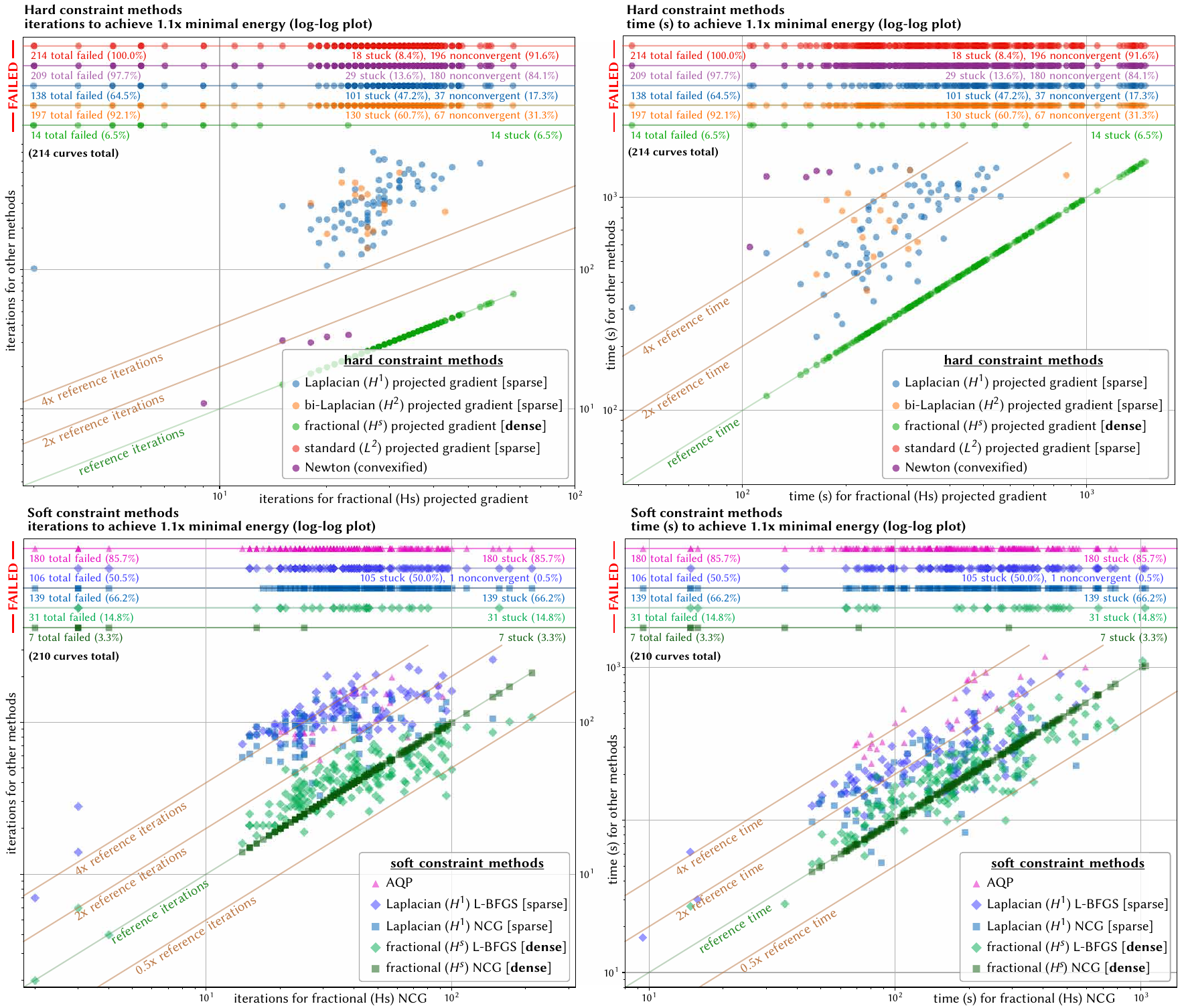}
   \caption{We used a dataset of about two hundred difficult knot embeddings to evaluate the performance of our strategy compared to the next most competitive methods.  Even without hierarchical acceleration, our fractional strategy was significantly faster---and succeeded at untangling a much larger fraction of knots.  Here we plot the time it took for each method to get within 1.1x of the reference energy, against the time taken by our fractional strategy.  Results have been split into hard/soft constraint enforcement (top/bottom rows), and iteration count/wall clock time (left/right columns).  At the top of each plot we show the number of failures after 24 minutes of compute time---\emph{stuck} indicates a failure of line search to make progress due to collisions; \emph{nonconvergent} means the method failed to get below 1.1x of the reference energy.\label{fig:ScatterPlots}}
\end{figure*}

\subsection{Local minimizers} As seen in \figref{LocalMinima}, the local minimizers found via our fractional descent strategy generally appear to be the same as with other schemes, up to rigid motions.  Hundreds more such examples can be found in the supplemental material.  Very rarely, two different methods produced local minimizers that were identical up to a \emph{reflection}; such \emph{amphichiral} pairs exist in some knot classes~\cite{Liang:1994:AK}, but of course have the same energy.

\subsection{Scaling behavior} 
\begin{figure}
  \includegraphics[width=\columnwidth]{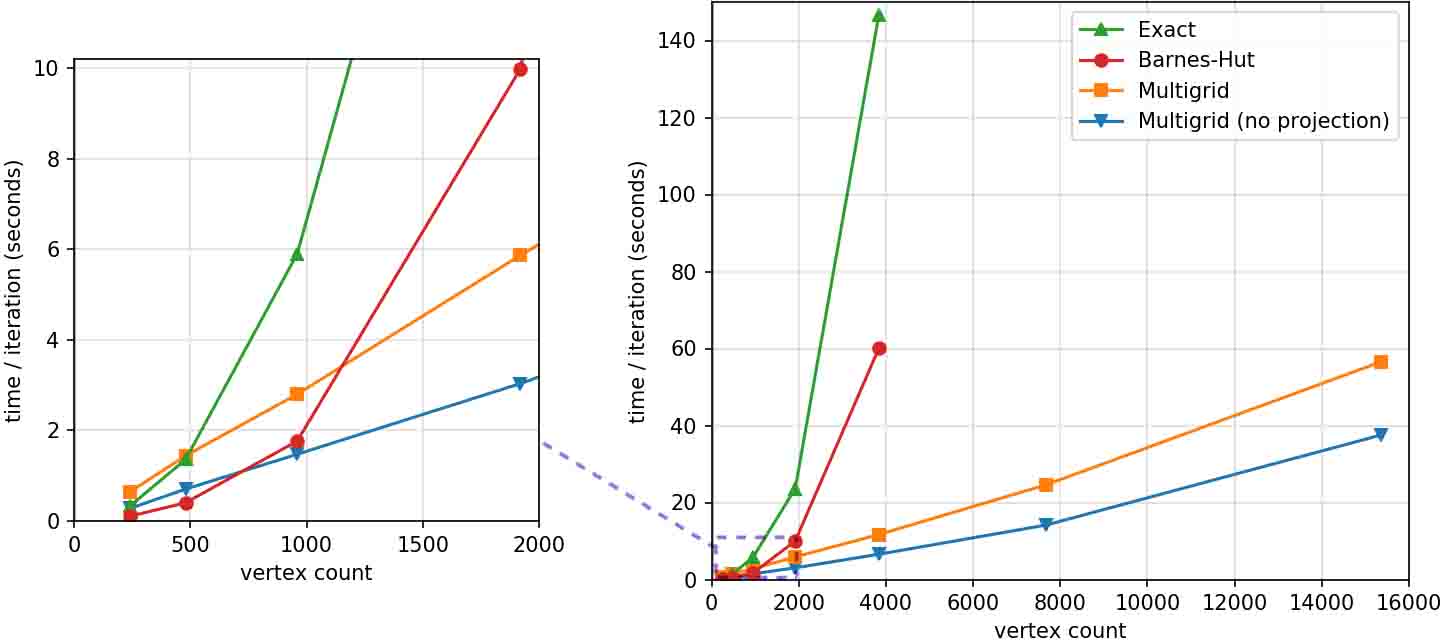}
   \caption{A comparison of runtime per iteration on samplings of the same curve with increasing resolution. ``Exact'' indicates no acceleration, ``Barnes-Hut'' indicates accelerated gradients only, and ``Multigrid'' indicates all accelerations enabled, with and without constraint projection. Reported numbers are averages over up to 500 iterations or until convergence.\label{fig:HierarchicalScaling}}
\end{figure}

We compared per-iteration costs of the unaccelerated scheme, a scheme using only Barnes-Hut (\secref{EnergyandDifferentialEvaluation}), and the full acceleration scheme described in \secref{Acceleration}---see \figref{HierarchicalScaling}. With full acceleration we observe near-linear scaling, whereas schemes that directly solve the dense system exhibit super-quadratic scaling and quickly run out of memory.  Note that constraint projection with direct solvers comes nearly for free, since a factorization of \eqref{ConstraintSaddlePointProblem} can be reused to solve \eqref{GradientSaddlePointProblem}.  In contrast, no reuse is possible in the fully accelerated scheme, making constraint projection relatively expensive. Disabling this step further speeds up the accelerated scheme, but leads to constraint drift over time.  Alternative methods for constraint enforcement (such as soft penalties, as noted above) might hence provide further improvement.

\section{Results and Applications}
\label{sec:ResultsAndApplications}

Given how ubiquitous plane and space curves are in areas like geometry, graphics, robotics, and visualization---and how natural it is to want to avoid collision of such curves---our method provides a useful computational framework for a wide variety of tasks.  Here we explore some preliminary applications that we hope will inspire future work.  All other examples in this section completed within a few minutes, except for the 3D curve packing example where we allowed curves to grow longer for several hours as a stress test.  We first describe constraints and potentials used for these examples.

\subsection{Constraints and Potentials}
\label{sec:ConstraintsAndPotentials}

A key feature of our optimization framework is that it not only efficiently minimizes knot energies, but that it can do so in conjunction with fairly arbitrary user-defined constraints and penalties (\secref{Constraints}).  This opens the door to a rich variety of computational design applications beyond the basic ``knot untangling'' that has been the focus of previous work.  For the applications that will be explored in \secref{ResultsAndApplications}, we consider the following constraints:

\begin{itemize}
   \item \textbf{Barycenter.} This fixes the barycenter of the curve to a point \(x_0\) via the constraint \(\Phi_{\mathrm{barycenter}}(\gamma) := \sum_{I \in E} \ell_I (x_I-x_0)\). In the absence of other constraints, this eliminates the null space of globally constant functions discussed in \appref{SobolevSlobodeckijGradient}.
   \item \textbf{Length.} The repulsive curve energy naturally wants to make the curve longer and longer. A simple way to counteract this is via a total length constraint \(\Phi_{\mathrm{length}}(\gamma) := L^0 - \sum_{I \in E} \ell_I\), where \(L^0\) is the target length.
   \item \textbf{Edge Length.} We can also constrain the lengths of each individual edge, allowing only isometric motions. This entails a constraint \(\Phi_{\mathrm{length},I}(\gamma) := \ell^0_I - \ell_I\) for each edge \(I\), where \(\ell^0_I\) is the target edge length.
   \item \textbf{Point Constraint.} To fix the position of a vertex \(i\) to the point \(x_i \in \RR^3\), we can add the constraint \(\Phi_{\mathrm{point},i}(\gamma) := \gamma_i - x_i\).
   \item \textbf{Surface Constraint.} To keep a point of the curve constrained to an implicit surface \(f(x) = 0\), we can add the constraint \(\Phi_{\mathrm{surface},i}(\gamma) := f(\gamma_i).\)
   \item \textbf{Tangent Constraint.} We can force the tangent \(T_I\) of an edge \(I\) to match a unit vector \(X \in \RR^3\) via the constraint \(\Phi_{\mathrm{tangent},I}(\gamma) := T_I - X\).
\end{itemize}

In several applications, we progressively increase or decrease the target length values $L_0$ or $l^0_I$; the next constraint projection step then enforces the new length.  We also consider the following penalties:

\begin{itemize}
   \item \textbf{Total length.} A simple energy is the total curve length, which provides a ``soft'' version of the total length constraint.  Discretely, this energy is given by \(\hEcal_{\text{length}}(\gamma) := \sum_{I \in E} \ell_I\).
   \item \textbf{Length difference.} This energy penalizes differences in adjacent edge lengths, given by \(\hEcal_{\text{diff}}(\gamma) = \sum_{v \in V_{\text{int}}} (\ell_{I_v} - \ell_{J_v})^2\), where $V_\text{int}$ denotes the set of ``interior'' vertices with degree 2, and $I_v$ and $J_v$ are the indicent edges to $v$.
   \item \textbf{Surface potential.} Given a surface \(M \subset \RR^3\), we use the energy \(\Ecal_M(\gamma) := \int_\gamma \int_M 1/|x_M - \gamma(x_\gamma)|^{\beta - \alpha} dx_M\ dx_\gamma\) to avoid collisions. This is effectively a Coulomb potential of the same order as \(\Energy\) on \(M\). In the discrete setting, \(M\) is a triangulated surface, and we use a BVH on \(M\) to accelerate the evaluation of \(\Ecal_M\) and its differential, in a similar fashion to \(\Energy\).
   \item \textbf{Field potential.} Given a fixed unit vector field \(X\) on \(\RR^3\), the energy \(\Ecal_X(\gamma) := \int_0^L |  T(x) \times X(\gamma(x)) |^2\ dx_\gamma\) encourages \(\gamma\) to run parallel (or anti-parallel) to \(X\).  We discretize this as \(\hEcal_X(\gamma) := \sum_{I \in E} \ell_I |\Tangent_I \times X(\Center_I)|^2\).
\end{itemize}

Note that the energies considered here involve lower-order derivatives than those in \(\Energy\), and do not therefore have a major effect on the stiffness of the overall system. Hence, we can continue to use the fractional Sobolev inner product without modification to define an efficient gradient flow.

\begin{figure}
   \includegraphics[width=\columnwidth]{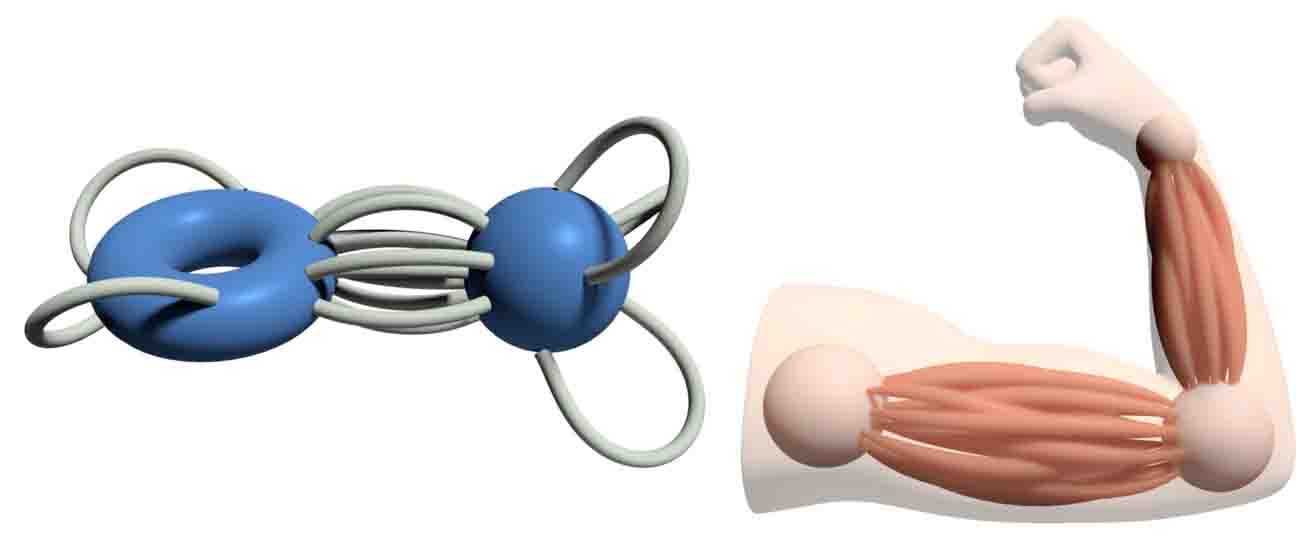}
   \caption{Allowing curves to slide freely over constraint surfaces \figloc{(left)} enables design tasks like arranging networks of muscles or muscle fibers \figloc{(right)}.\label{fig:PartialSurfaceConstraint}}
\end{figure}

\begin{figure}
   \includegraphics[width=\columnwidth]{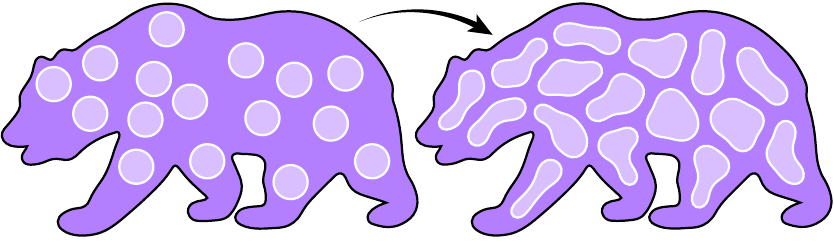}
   \caption{Just as repulsive potentials are commonly used to find equally-distributed points, we can compute collections of equally-spaced curves (here constrained to a region via a fixed curve potential).\label{fig:BlueNoise}}
\end{figure}

\subsection{Curve Packing}
\label{sec:CurvePacking}

\begin{figure}
   \includegraphics[width=\columnwidth]{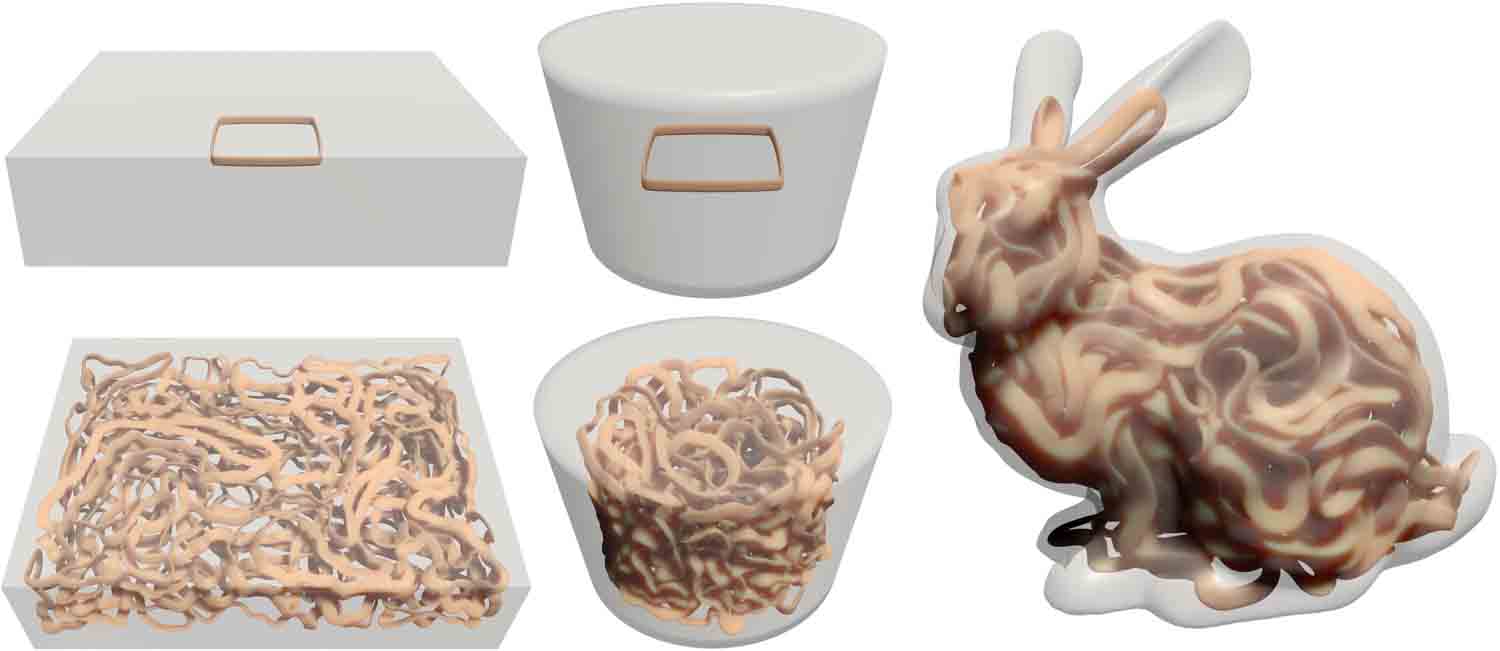}
   \caption{By penalizing proximity to a fixed surface, we can pack curves into any domain.  Progressively increasing edge length forces curves to maintain a balance between surface avoidance and self-avoidance. (Here we render curves with a non-circular cross section, which is not modeled by the energy.)\label{fig:SurfacePenalty}}
\end{figure}

\begin{figure}
   \includegraphics[width=\columnwidth]{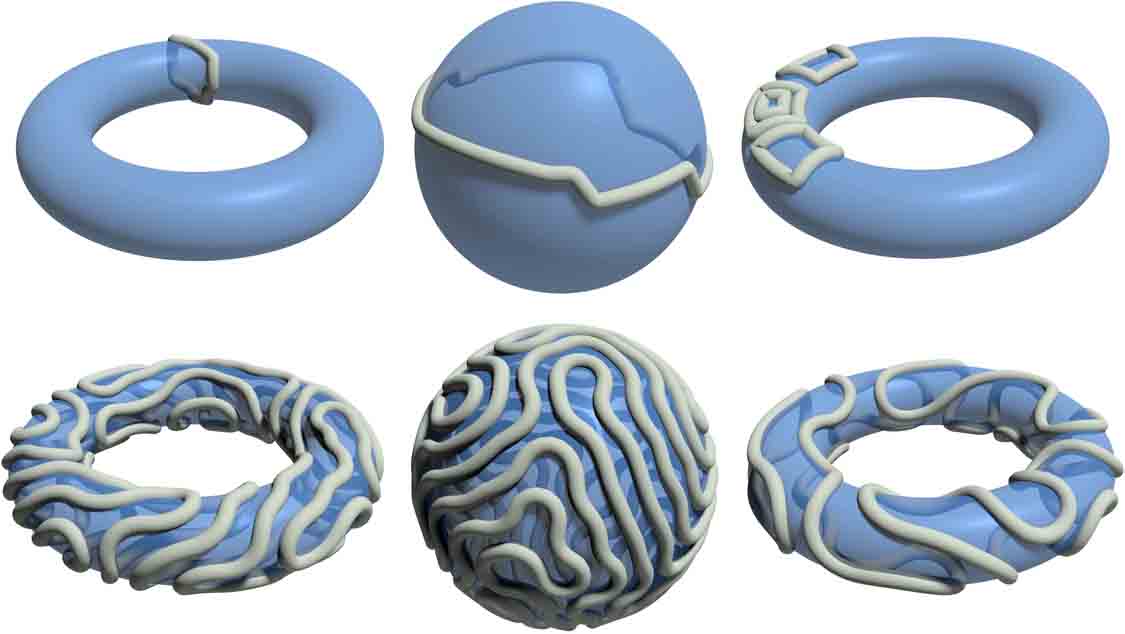}
   \caption{Patterns obtained by constraining a collection of repulsive curves to a surface and increasing their lengths (initial states shown above their final configurations).\label{fig:SurfaceConstraint}}
\end{figure}

\begin{figure}
   \includegraphics[width=\columnwidth]{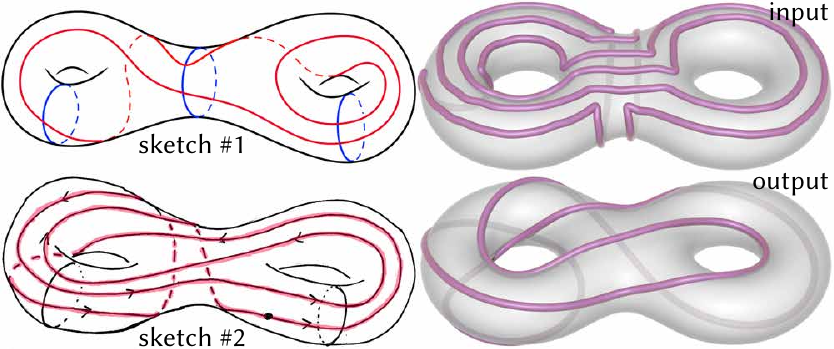}
   \caption{Loops arising in topology can be difficult to draw by hand---the sketches at left were done by Nathan Dunfield to illustrate \emph{Dehn-Thurston} coordinates.  At right we generate an equispaced version of this curve by flowing a rough sketch, subject to an implicit surface constraint.\label{fig:TorusLoops}}
\end{figure}

Packing problems (such as \emph{bin packing}) appear throughout geometry and computer graphics, playing an important role in, \eg{}, 2D layouts for manufacturing or UV atlas generation.  An adjacent problem is generation of regular sampling patterns, \eg{}, blue noise sampling via Poisson disk rejection.  The ability to optimize large families of repulsive curves enables us to solve analogous ``curve packing'' problems---for instance, in \figref{BlueNoise}, we use a fixed boundary curve to pack disks of increasing length; likewise, in \figrefs{BunnyTeaser,SurfacePenalty}, we use a surface penalty to pack increasingly long curves into a target region.  \figref{SurfaceConstraint} likewise packs increasingly long curves on a surface. Going the opposite direction, we can also \emph{decrease} length while encouraging repulsion to generate clean illustrations that are difficult to draw by hand (\figref{TorusLoops}). Finally, by constraining only parts of curves to lie on surfaces, we can design biologically-inspired curve networks such as muscle fibers (\figref{PartialSurfaceConstraint}), which are attached to objects at their endpoints but are otherwise free.

\subsection{Graph Drawing}
\label{sec:GraphDrawing}

\begin{figure}
   \centering
   \includegraphics{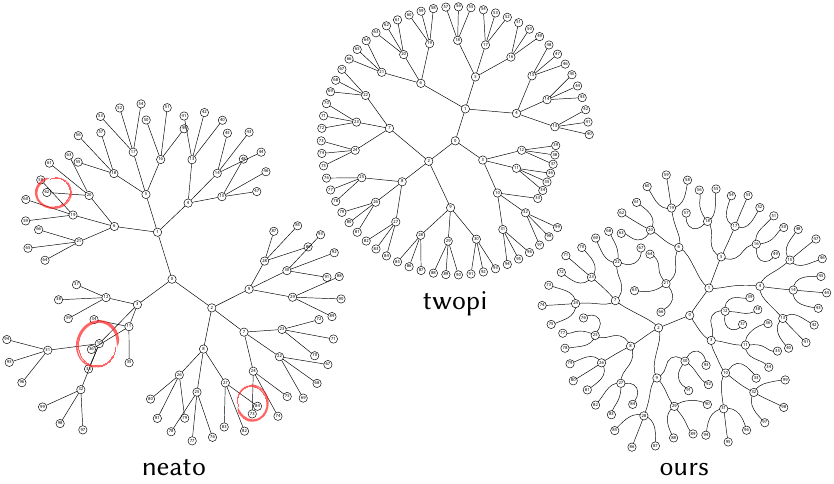}
   \caption{Traditional 2D graph drawing algorithms based on nodal proximity may cause edges to cross \figloc{(left)} or position nodes extremely close together \figloc{(center)}; these layouts were produced by the popular \emph{Graphviz} library~\cite{Ellson:2001:GVO}. By treating edges as repulsive curves, we can obtain graph drawings that are both more compact and more legible \figloc{(right)}.\label{fig:GraphDrawing2D}}
\end{figure}

\begin{figure}
   \includegraphics{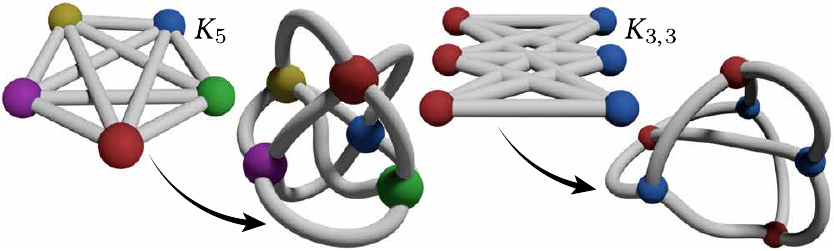}
   \caption{Isometric embedding: by jittering 2D drawings of non-planar graphs (which necessarily have crossings), curve repulsion with length constraints yields nicely spaced embeddings in \(R^3\) with prescribed edge lengths.\label{fig:GraphDrawing3D}}
\end{figure}

\begin{figure}
   \includegraphics[width=\columnwidth]{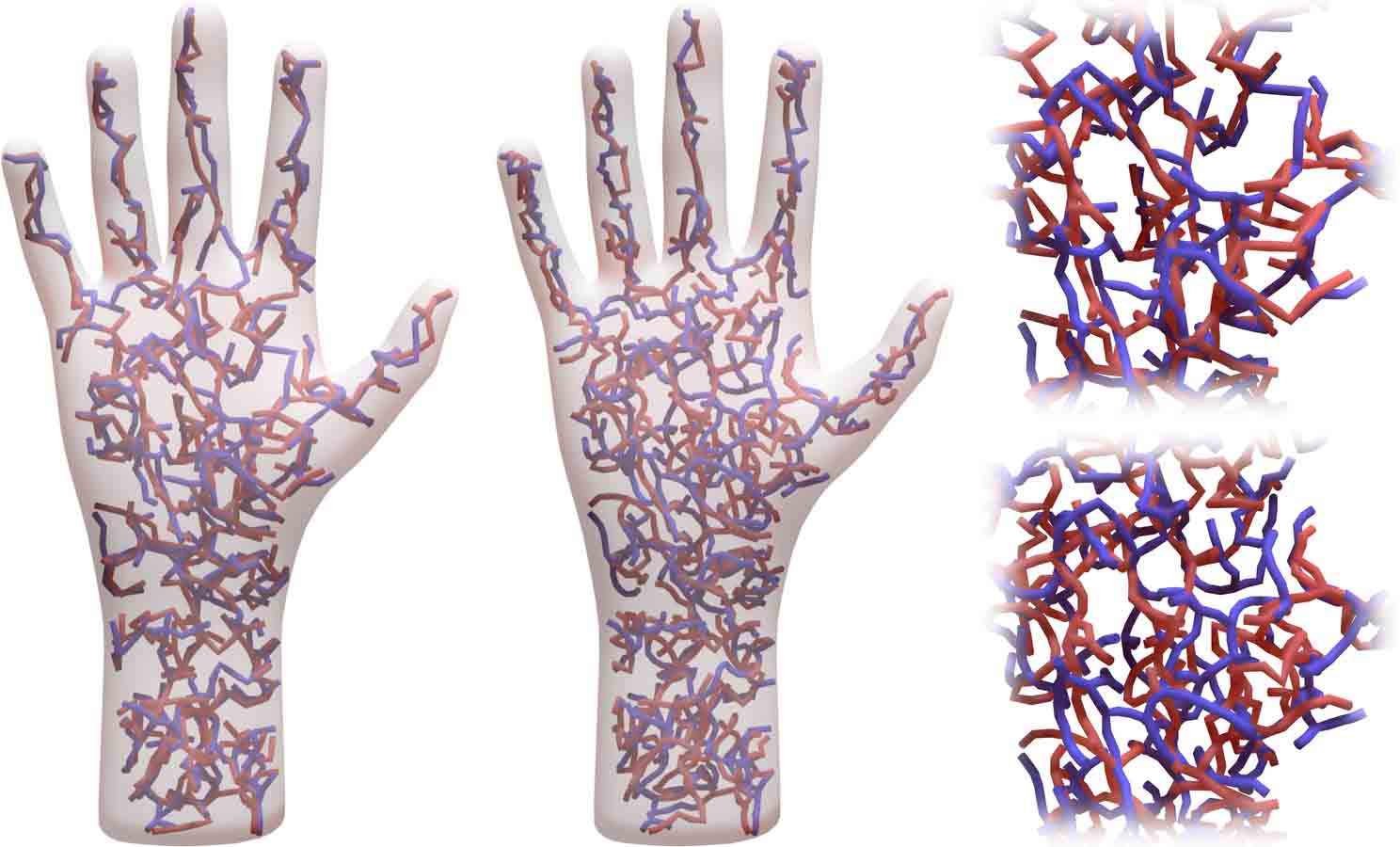}
   \caption{A fast solver for curves facilitates the design of complex curve networks such as this pair of interwoven vascular networks. Starting from a crude initial topology \figloc{(left)}, curve repulsion produces a network with the same endpoints, but improved smoothness and spacing \figloc{(center, right)}.\label{fig:CapillaryNetwork}}
\end{figure}

A basic problem in data visualization is drawing \emph{graphs}; a typical approach is to use a force-based layout that seeks to avoid, \eg{}, collisions between nodes, or over/under-extension of edges~\cite{Fruchterman:1991:GDF}.  Our framework makes it easy to optimize the geometry of the edges themselves, opening the door to graph layouts that are both more compact and more legible (\figref{GraphDrawing2D}).  We can also use this machinery to obtain legible drawings of nonplanar graphs, by perturbing a planar embedding (\figref{GraphDrawing3D}); here, the ability to preserve lengths conveys information about edge weights.  A particularly interesting graph embedding problem is the design of synthetic hydrogel \emph{vascular networks}~\cite{Grigoryan:2019:MVN}; \figref{CapillaryNetwork} shows a simple example where we optimize a multivascular network (starting from subgraphs of a tet mesh and its dual).

\setlength{\columnsep}{1em}
\setlength{\intextsep}{0em}
\begin{wrapfigure}{r}{67pt}
   \includegraphics{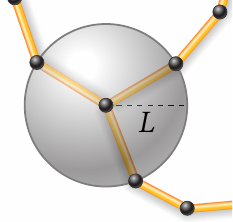}
\end{wrapfigure}
Note that at junctures between more than two edges, the tangent-point energy will always be large (since three or more edges cannot be collinear), rapidly forcing vertices away from each other. This can be counteracted by constraining their edge lengths, forcing the vertices to lie on spheres of constant radii around the junctures.

\subsection{Self-Avoiding Splines}
\label{sec:Self-AvoidingSplines}

\begin{figure}[b]
   \includegraphics[width=.8\columnwidth]{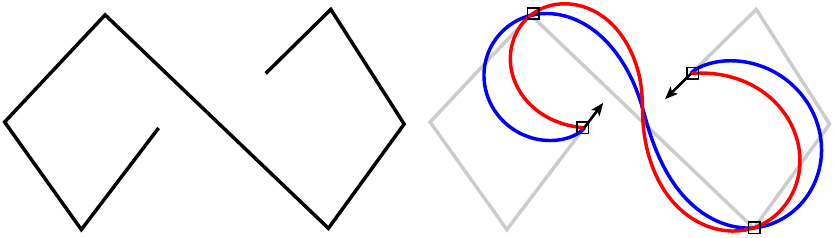}
   \caption{As with B\'{e}zier curves, we can also control curve tangents at both interior and endpoints.  Here we flow a polygonal curve \figloc{(left)}, to a smooth interpolant with fixed points \figloc{(red)}, and fixed points and tangents \figloc{(blue)}. \label{fig:TangentInterpolation}}
\end{figure}

\begin{figure}
   \includegraphics[width=\columnwidth]{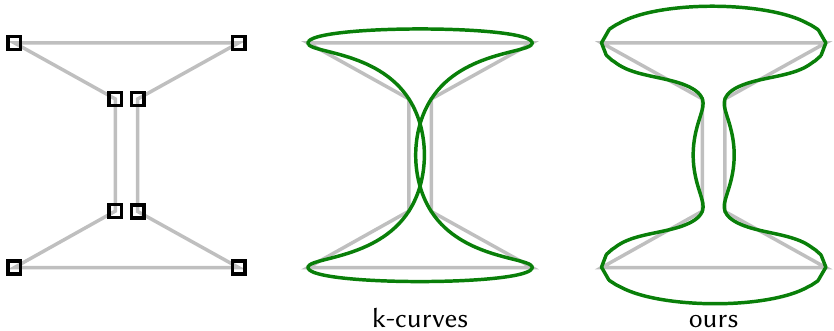}
   \caption{Standard curve interpolation methods in 2D drawing programs can cause curves to self-intersect \figloc{(center)}, even when the control polygon \figloc{(left)} does not. By starting from the control polygon and constraining the control points, we obtain a smooth, non-intersecting interpolant \figloc{(right)}.\label{fig:CurveInterpolation}}
\end{figure}

Beyond standard B\'{e}zier input, sophisticated tools have been developed for drawing spline curves---but do not consider the basic constraint of ensuring that curves do not cross themselves (which is often desirable for physical or aesthetic reasons).  For instance, \figref{CurveInterpolation} (\figloc{center}) shows the interpolation of a set of control points by \emph{k-curves}~\cite{Yan:2017:KCI}, which underpin one of the basic drawing tools in Adobe Illustrator (the \emph{Curvature Tool}).  By simply applying point constraints at the control points, and letting the length increase under our repulsive flow, we obtain a nice interpolating curve without self-intersection (\figref{CurveInterpolation}, \figloc{right}).  In this context we can also use our tangent constraint to control the behavior of such a curve at open endpoints (\figref{TangentInterpolation}).

\subsection{Multi-agent Path Planning}
\label{sec:Multi-AgentPathPlanning}

\begin{figure}[b]
   \includegraphics[width=\columnwidth]{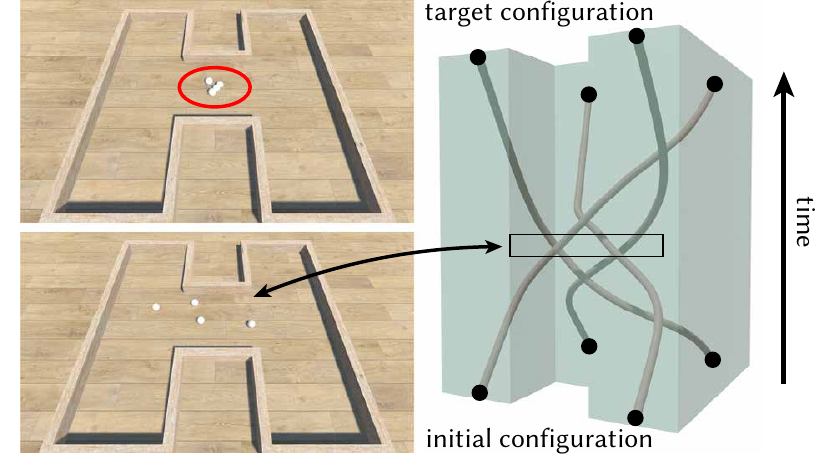}
   \caption{\figloc{Top-left:} In this path planning scenario, an initial trajectory brings the four agents dangerously close together. \figloc{Bottom-left:} By treating trajectories as curves in space-time, our system provides solutions that maximally avoid collisions, making them more robust to control errors. \figloc{Right:} Finding 2D trajectories is equivalent to optimizing a 3D braid with fixed endpoints constrained to an extrusion of the given environment.  This same construction can easily be generalized to 3D environments.\label{fig:PathPlanning}}
\end{figure}

In robotics, numerous algorithms have been developed for the problem of multi-agent path planning~\cite{deWilde:2013:PRC}, wherein multiple agents must travel from fixed start to end locations without colliding with the environment or each other. Many algorithms operate on a discrete grid or graph~\cite{Yu:2013:MAP}, which quantizes the solution space and does not penalize near-collisions; such trajectories may therefore not be robust to sensing or control error. By treating path planning as a space-time optimization of continuous curves with fixed endpoints, we can use curve repulsion to find (or refine) trajectories that maximize collision avoidance, making them more resilient to error (\figref{PathPlanning}). Finding such trajectories in \(n\) dimensions is equivalent to optimizing a braid in \(n+1\) dimensions; since neither the size of the curve nor the cost of a BVH/BCT depends strongly on dimension, this strategy easily generalizes to three (or more) dimensions.

\begin{figure}
   \includegraphics[width=\columnwidth]{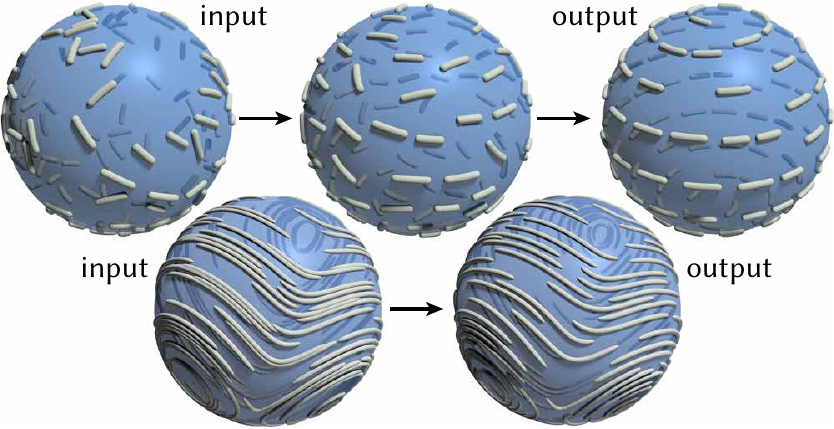}
   \caption{Encouraging curve tangents to align with a given vector field improve the quality of streamline visualization. Here, a random set of curve segments \figloc{(top)} aligns itself with a rotational vector field; we can also optimize randomly sampled streamlines \figloc{(bottom)} to improve their spacing.\label{fig:FlowVisualization}}
\end{figure}

\subsection{Streamline Visualization}
\label{StreamlineVisualization}

A common way to visualize vector fields is by tracing integral curves or \emph{streamlines}; significant effort has gone into algorithms that provide uniform spacing (\eg{}, by incrementally constructing a Delaunay triangulation~\cite{Mebarki:2005:FPS}), though such methods can be difficult to generalize to 3D volumes or vector fields on surfaces.  We can generate nicely-spaced streamlines by adding a field alignment potential to the tangent-point energy---for instance, in \figref{FlowVisualization} we start with a set of random curve segments, which automatically coalesce into streamlines.

\section{Limitations and Conclusion}
\label{sec:LimitationsandConclusion}

\setlength{\columnsep}{1em}
\setlength{\intextsep}{0em}
\begin{wrapfigure}{r}{100pt}
   \includegraphics{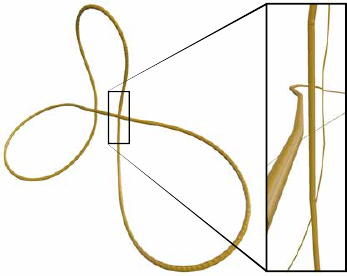}
\end{wrapfigure}
Since we approximate the tangent-point energy via numerical quadrature, it is possible for a very coarse curve to pass through the energy barrier.  However, crossings can be prevented via continuous time collision detection (\secref{TimeStepping}); to maintain accuracy one could also try adding more quadrature points at the previous time step if any collisions occur. For the design tasks in this paper, we did not find such strategies necessary.  Also on very coarse meshes, edges that are extremely close together can temporarily get stuck in a near-crossing configuration (see inset).  In this situation, the term \(\Kernel[2][4]\) from the low-order term (\eqref{LowOrderTerm}) is very large, causing the inverse of \(A\)---and hence the Sobolev gradient---to be very small.  One idea is to use adaptive quadrature for edge pairs that are close in space, which would better resolve the near-infinite high-order term and hence push the curve apart.  Given the scalability of our approach, another pragmatic solution is simply to increase the overall resolution.

There are many ways to further accelerate our solver.  For instance, we did not vectorize our code, parallelized only the matrix-vector multiply in non-well-separated leaves of the BCT, and did not make use of the GPU.  For small time steps one might re-fit rather than re-build the BVH; likewise, it may be beneficial to incrementally update the BCT.  Better line search or descent direction heuristics may also reduce the overall number of steps.

Finally, a natural question is how to extend these techniques to \emph{surface} repulsion.  The tangent-point energy seems attractive here since (unlike M\"{o}bius energy) it needs only Euclidean rather than geodesic distances.  One now has double integrals over \emph{surfaces}, but might still achieve efficiency via hierarchical acceleration.  In general, we are hopeful our investigation will provide valuable insight into using repulsive energies for computational design.

\begin{figure}
   \includegraphics[width=\columnwidth]{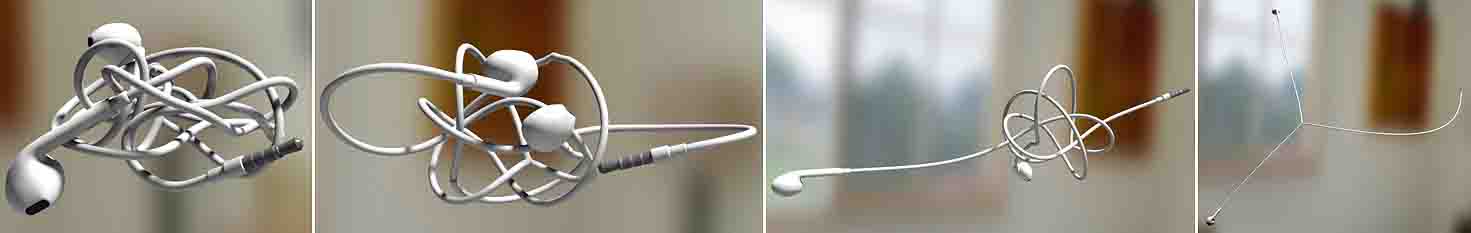}
   \caption{Untangling a pair of earbuds via repulsion (see supplemental video).\label{fig:CableUntangle}}\vspace{-\baselineskip}
\end{figure}

\begin{acks}
   Thanks to Stelian Coros for early discussion of these ideas.  The bunny mesh is used courtesy of the Stanford Computer Graphics Laboratory.  This work was supported by a Packard Fellowship, NSF awards 1717320 and 1943123, and gifts from Autodesk, Adobe, Activision Blizzard, Disney, and Facebook.  The second author was supported by a postdoc fellowship of the German Academic Exchange Service and by DFG-Project 282535003: \emph{Geometric curvature functionals: energy landscape and discrete methods}.  The third author was also supported by NSF award DMS-1439786 and Sloan award G-2019-11406 while in residence at ICERM.
\end{acks}

\bibliographystyle{ACM-Reference-Format}
\bibliography{SelfAvoiding}

\appendix

\section{Sobolev-Slobodeckij Gradient}
\label{app:SobolevSlobodeckijGradient}

How do we obtain an ideal gradient flow for the tangent-point energy (\ie, one that behaves like an ODE)?  Unlike standard energies (elastic energy, Willmore energy, \etc{}), an answer to this question has not yet been worked out formally.  However, we can make an educated guess based on past wisdom about curve energies.

In general, suppose an energy \(\mathcal{E}\) has a (Fr\'{e}chet) differential \(d\mathcal{E}\).  To determine the highest-order derivatives, it is not necessary to derive an explicit expression for \(d\mathcal{E}\) as we did for the Dirichlet energy (\secref{WarmUpDirichletEnergy}).  Instead, we can reason about the associated function spaces:  as long as we know the \emph{order} of \(d\mathcal{E}\), we can ``cancel'' spatial derivatives by constructing an inner product of the same order.

For the tangent-point energy, existing analysis gives the maximum order of derivatives in \(\Energy\) (\appref{EnergySpace}), from which we deduce the order of \(d\Energy\) (\appref{OrderOfTheDifferential}).  What is unusual here is that the number of derivatives is \emph{fractional} (\appref{FractionalSobolevSpaces}); to build an inner product of appropriate order, we therefore start with the \emph{fractional Laplacian} (\secref{FractionalAnalysis}), and formulate an analogous operator for embedded curves.  Taking further (integer) derivatives then yields an operator of the same order as \(d\Energy\) (\appref{FractionalInnerProduct}).  From there, we use additional heuristics (inspired by numerical experiments) to choose a low-order term that makes this operator well-behaved and invertible (\appref{LowOrderTerm}), allowing us to use it in the definition of a fractional Sobolev gradient (\secref{FractionalSobolevGradient}).

\subsection{Fractional Analysis}
\label{sec:FractionalAnalysis}

We begin with a brief discussion of Sobolev spaces of \emph{fractional} order \(k \notin \ZZ\); for further background, see~\cite{DiNezza:2012:HGF}.

\subsubsection{Fractional Differential Operators}
\label{sec:FractionalDifferentialOperators}

\setlength{\columnsep}{1em}
\setlength{\intextsep}{0em}
\begin{wrapfigure}{r}{95pt}
   \includegraphics{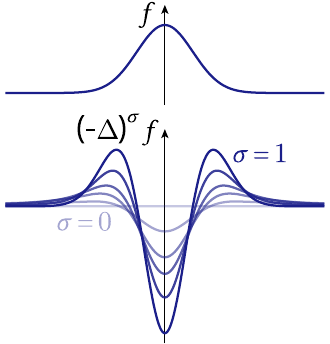}\vspace{-.5\baselineskip}
   \caption{Fractional Laplacian of \(f\) for several values of \(\sigma\).\label{fig:FractionalLaplacian}}
\end{wrapfigure}
Whereas standard differential operators \(L\) are purely \emph{local} (\ie, the value of \((Lu)(x)\) depends only on an arbitrarily small neighborhood of \(u(x)\)), fractional differential operators are \emph{nonlocal} (\((Lu)(x)\) can depend on the value of \(u\) at any point \(y\)).  Since the tangent-point energy is nonlocal, it will also have nonlocal derivatives.  Hence, finding an inner product well-matched to its gradient flow entails constructing an appropriate fractional differential operator---an important example in our setting is the \emph{fractional Laplacian} \((-\Delta)^\sigma\) on \(\RR^\DomDim\), which is commonly defined by taking powers of the eigenvalues in the spectral expansion.  For \(\sigma \in (0,1)\) and all sufficiently regular \(u,v: \RR^\DomDim \to \RR\), the operator can also be expressed via the integral
\begin{equation}
   \label{eq:FractionalLaplacian}
   \llangle (-\Delta)^\sigma u, v \rrangle = C \iint_{\RR^\DomDim \times \RR^\DomDim} \!\!
 		\frac{u(x)\!-\!u(y)}{|x-y|^{\sigma}} \, \frac{v(x)\!-\!v(y)}{|x-y|^{\sigma}} \, \frac{dx dy}{|x-y|^{\DomDim}},
\end{equation}
where the constant \(C \in \RR\) depends only on \(\DomDim\) and \(\sigma\) \cite{Kwasnicki:2017:TED}.  The behavior of this operator is illustrated in \figref{FractionalLaplacian}. 

\subsubsection{Fractional Sobolev Spaces}
\label{app:FractionalSobolevSpaces}

\begin{figure}[b]
   \includegraphics[width=\columnwidth]{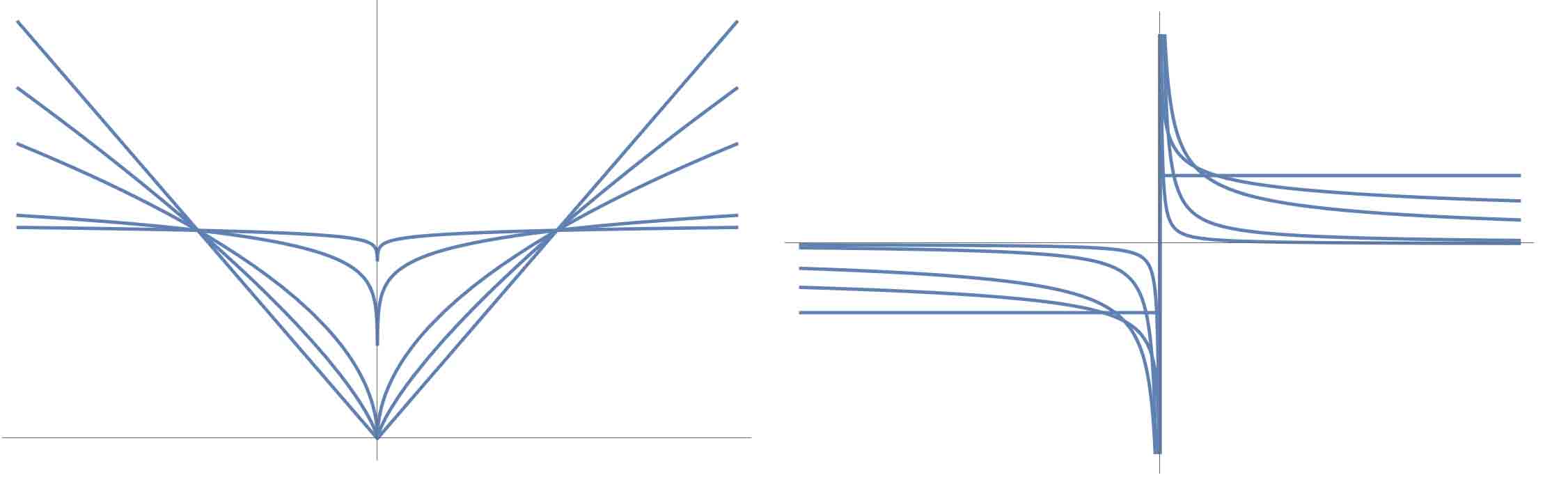}
   \caption{The curves \((x,|x|^\sigma)\) are examples of curves in \(W^{\sigma,p}\) \figloc{(left)}.  Their 1st derivatives are not \(L^p\) integrable \figloc{(right)}.\label{fig:FractionalCurve}}
\end{figure}

There are two common ways to understand Sobolev spaces of fractional order.  One is to consider the Fourier transform of the Laplacian \(\Delta\), leading to the \emph{Bessel potential spaces}~\(H^{s,p} := (-\Delta)^{-s/2} (L^p)\) \cite[Section 2.2.2]{TriebelTFS1983}. For us, however, this viewpoint helps only to understand the case \(W^{s,2}\).  The other, essential for studying the tangent-point energy, is via the \emph{Sobolev-Slobodeckij spaces} \(W^{k+\sigma,p}\).  Functions \(u\) in these spaces look like functions in an ordinary Sobolev space, but with a nonlocal regularity condition on the highest-order derivative \(u^{(k)}\).  In particular, suppose we write \(s = k+\sigma\) for \(k \in \mathbb{Z}_{\geq 0}\) and \(\sigma \in (0,1)\). 
Then, on an $\DomDim$-dimensional Riemannian manifold \(\Domain\), one defines
\[
   W^{k+\sigma,p}(\Domain) := \big\{ u \in W^{k,p}(\Domain) \ \big|\ [u^{(k)}]_{W^{\sigma,p}} < \infty \big\}.
\]
The expression in square brackets is the (Gagliardo) semi-norm
\[
   [u]_{W^{\sigma,p}} := \bigg( \iint_{\Domain^2} \left| \frac{u(x)\!-\!u(y)}{d(x,y)^{\sigma}} \right|^p\ \frac{dx\ dy}{d(x,y)^\DomDim} \bigg)^{1/p},
\]
where \(d(x,y)\) is the shortest distance between \(x\) and \(y\) in \(\Domain\).  Just as a Lipschitz function is more regular than an arbitrary continuous function without being differentiable, a function in \(W^{k+\sigma,p}\) is more regular than one in \(W^{k,p}\), without getting a whole additional derivative (\ie, \(W^{k+1,p} \subsetneq W^{k+\sigma,p}\)).  \figref{FractionalCurve} shows an example.

\paragraph{Dual Space.} Just as the dual of the classical Sobolev space \(W^{k,p}\) is \(W^{-k,q}\) (where \(1/p + 1/q = 1\)), the dual of the Sobolev-Solobdeckij space \(W^{s,p}\) can be characterized as a space with ``\(-s\) derivatives'' in the sense that the fractional Laplacian \( (-\Delta)^s\) identifies \(W^{s,p}\) with \(W^{-s,q} := (W^{s,p})^*\)~\cite[Remark 2.5]{DiNezza:2012:HGF}.

\subsection{Energy Space}
\label{app:EnergySpace}

To determine the order of the tangent-point differential \(d\Energy\), we first consider the biggest space of functions for which the energy \(\Energy\) is well-defined.  \citet{Blatt:2013:EST} gives the following condition on the differentiability of the curve \(\gamma\) (see also \citet{Blatt:2015:RTT}):
\begin{lemma}
   \label{lem:FiniteTangentPointEnergy}
   Suppose $\NumExp > 1$ and $\DenomExp \in [\NumExp + 2, 2\,\NumExp + 1)$, let $s \ceq \frac{\DenomExp}{\NumExp} -\frac{1}{\NumExp}$, and consider an embedded curve \(\gamma \in C^1(S^1;\RR^3)\).  Then \(\gamma\) has finite tangent point energy \(\Energy[\NumExp][\DenomExp](\gamma)\) if and only if, up to reparameterization, \(\gamma \in \Sobo[s,\NumExp][][S^1][\RR^3]\).
\end{lemma}
In other words, the tangent point energy is well-defined only for curves that have an \(s\)th derivative, and for which the \(\NumExp\)th power of that derivative is integrable---for example, it will not be finite for a polygonal curve.  The somewhat unusual situation is that \(s\) is not an integer: instead, it is a fractional value in the interval \((1,2)\).

\subsection{Order of the Differential}
\label{app:OrderOfTheDifferential}

In general, if an energy \(\mathcal{E}\) is defined for functions in a space \(X\), then its differential \(d\Energy\) will have the prototype \(d\mathcal{E}: X \to X^*\), where \(X^*\) is the dual space.  For instance, the Dirichlet energy \(\DirichletEnergy\) operates only on functions \(f \in H^1\).  Hence, its differential is a map \(d\DirichletEnergy: H^1 \to (H^1)^*\), which we saw explicitly in \secref{WarmUpDirichletEnergy}: given a function \(f \in H^1\), \(d\DirichletEnergy|_f\) produces a linear map \(\llangle -\Delta f, \cdot \rrangle\) from functions in \(H^1\) to real numbers, \ie, an element of \((H^1)^*\).

In the case of the tangent point energy, then, we get that \(d\Energy\) is a map from \(W^{s,p}\) to the dual space \((W^{s,p})^* = W^{-s,q}\) (\secref{FractionalDifferentialOperators}).  Hence, \(d\Energy\) is a ``differential operator'' of order \(2s\), \ie, it reduces the differentiability of its argument by \(2s\).  To get a well-behaved flow, we should therefore pick an inner product of the same order, and (for computational purposes) is reasonably easy to invert.

\subsection{Fractional Inner Product}
\label{app:FractionalInnerProduct}

Just as one uses the Laplace operator \(\Delta\) to define integer Sobolev inner products, we use a fractional operator to define a fractional Sobolev inner product.  For an embedded curve \(\gamma: \Domain \to \RR^3\), one idea is to start with the 1D fractional Laplacian \((-\Delta)^\sigma\).  Alternatively, we can define an analogous operator by replacing the intrinsic distance \(|x-y|\) on the right-hand side of \eqref{FractionalLaplacian} with the extrinsic distance \(|\gamma(x)-\gamma(y)|\) between points in the embedding. This latter construction yields an operator \(L_\sigma\) defined by the relationship
\begin{equation}
   \label{eq:FractionalPseudoLaplacian}
   \llangle L_\sigma u, v \rrangle 
   := 
   \!\!
   \iint_{\Domain^2}  \!\!
   \frac{u(x)\!-\!u(y)^{\phantom{\sigma}}}{|\gamma(x)\!-\!\gamma(y)|^\sigma} \, 
   \frac{v(x)\!-\!v(y)^{\phantom{\sigma}}}{|\gamma(x)\!-\!\gamma(y)|^\sigma} \, \frac{dx_\gamma dy_\gamma}{|\gamma(x)\!-\!\gamma(y)|}
\end{equation}
for all sufficiently regular \(u,v: \Domain \to \RR\).  For any \(\sigma \in (0,1)\), both \((-\Delta)^\sigma\) and \(L_\sigma\) are fractional operators of order \(2\sigma\).  But the benefit of \(L_\sigma\) is that it requires only Euclidean distances---which for embedded curves are easier to evaluate than geodesic distances.  Moreover, building a fractional Laplacian via an explicit Fourier transform is prohibitively expensive, requiring a full eigendecomposition of a discrete Laplace matrix.  In contrast, integral expressions like Equations \ref{eq:FractionalLaplacian} and \ref{eq:FractionalPseudoLaplacian} can easily be evaluated \ala{} \secref{DiscreteLowOrderTerm}, and accelerated using hierarchical techniques \ala{} \secref{Acceleration}.

\subsubsection{High-Order Term}
\label{app:HighOrderTerm}

To get an inner product of the same order as \(d\Energy\), we compose the operator \(L_\sigma\) with further (integer) derivatives \(\cD\).  In particular, \lemref{FiniteTangentPointEnergy} implies that \(s = 1 + \sigma\) for \(\sigma \in (0,1)\).  Hence, to define an operator \(B_\sigma\) of order \(2s = 2\sigma + 2\), we apply two additional derivatives to \(L_\sigma\), \ie, we say that
\[
   \llangle B_\sigma u, v \rrangle := \llangle L_\sigma \cD u, \cD v \rrangle
\]
for all sufficiently regular \(u,v: \Domain \to \RR\).  This relationship provides the definition of \(B_\sigma\) in \eqref{HighOrderTerm}.

\subsubsection{Low-Order Term}
\label{app:LowOrderTerm}

As discussed in \secref{HighOrderTerm}, the operator \(B_\sigma\) is not invertible.  We hence add the low-order term \(B^\low_\sigma\) from \eqref{LowOrderTerm}.  Since \(B_\sigma\) and \(B^\low_\sigma\) exhibit the same scaling behavior under a rescaling of \(\gamma\), the behavior of the resulting gradient flow will not depend on the global scale.  To see why, consider a rescaling of the curve \(\gamma \mapsto c\gamma\) by a factor \(c > 0\).  Then \(\cD\) scales by a factor \(1/c\), the term \(1/|\gamma(x)-\gamma(y)|^{2s+1}\) scales by \(1/c^{2s+1}\), and the measure \(dx_\gamma dy_\gamma\) scales by \(c^2\).  Then \(B_\sigma\) scales by \(c^2 / (c^2 c^{2s+1}) = 1/c^{2s+1},\) and \(L_\sigma\) scales by just \(c^2/c^{2s+1}\).  Hence, to get \(B^\low_\sigma\) we multiply \(L_\sigma\) by \(\Kernel[2][4]\), which scales like \(1/c^2\) (since it has \(c^2\) in the numerator, and \(c^4\) in the denominator).  More generally, one could use \(\Kernel\) for any \(\alpha,\beta\) such that \(\alpha - \beta = -2\).  This low-order term also tends to accelerate the evolution of the flow by preserving near-constant motions that slide near-tangentially and do not tend toward collision (\figref{L2HardCase}).

\section{Acceleration Scheme}
\label{app:AccelerationScheme}

\subsection{Energy and Differential Evaluation}
\label{app:EnergyAndDifferentialEvaluation}

\subsubsection{Bounding Volume Hierarchy}
\label{app:BoundingVolumeHierarchy}

To build the BVH we first construct tangent-points \(p_I := (\Tangent_I,\Center_I) \in \RR^6\) for each edge \(I \in E\).  We then cycle through all six coordinates, choosing a splitting plane that minimizes the sum of squared diameters of the two child bounding boxes.  Below a user-specified threshold, all remaining tangent-points are placed in a single leaf node.   In each node \(\Node\) we also store data needed for Barnes-Hut.  Specifically,
\[
      \NodeMass_\Node    \!:=\!\! \sum_{I \in \Node} \ell_I, \qquad
      \NodeCenter_\Node  \!:=\!\! \sum_{I \in \Node} \ell_I \Center_I / \NodeMass_\Node, \qquad
      \NodeTangent_\Node \!:=\!\! \sum_{I \in \Node} \ell_I \Tangent_I / \NodeMass_\Node,
\]
give the total mass, center of mass, and (length-weighted) average tangent, \resp{}; we will use \(\overline{p}_\Node := (\NodeTangent_\Node,\NodeCenter_\Node)\) to denote the corresponding tangent-point.  We also store the bounding box radii \(r_\Center^\Node\) and \(r_\Tangent^\Node\) with respect to spatial and tangential coordinates, \resp{}

\subsubsection{Barnes-Hut Approximation}
\label{app:BarnesHutApproximation}

To evaluate the energy for a tangent-point \(p_\TargetEdge = (\Tangent_\TargetEdge,\Center_\TargetEdge) \in \RR^6\) with mass \(\Mass_\TargetEdge \in \RR\), we traverse the BVH from the root, checking at each node if a local approximation is \emph{admissable} (see below).  If so, we evaluate the approximation
\begin{equation}
   \label{eq:BVHNodeTPE}
   (\Energy)_{\TargetEdge \SourceNode} 
   := 
   \frac{
   	| \Tangent_\TargetEdge \times (\Center_\TargetEdge - \NodeCenter_\SourceNode) |^\NumExp
   }{
   	|\Center_\TargetEdge - \NodeCenter_\SourceNode|^\DenomExp
   } \Mass_\TargetEdge \NodeMass_\SourceNode.
\end{equation}
and terminate traversal; otherwise, we sum the energy of the two children.  If we reach a leaf node \(\SourceNode\), we directly add up the contributions of the edges contained in this node, \ie,
\[
   \sum_{\SourceEdge \in \SourceNode}
   \frac{
   	|\Tangent_\TargetEdge \times (\Center_\TargetEdge - \Center_\SourceEdge)|^\NumExp
   }{
   	|\Center_\TargetEdge - \Center_\SourceEdge|^\DenomExp
   }  \ell_\TargetEdge \ell_\SourceEdge.
\]

\paragraph{Admissibility} A simple Taylor series analysis of \eqref{BVHNodeTPE} indicates that to keep approximation error below a user-specified threshold \(\varepsilon > 0\), it is sufficient to ensure that
\begin{equation}
   \label{eq:BarnesHutCondition}
   r_\Center^\SourceNode/|\Center_\TargetEdge - \NodeCenter_\SourceNode| \lesssim \varepsilon \qquad \text{and} \qquad r_\Tangent^\SourceNode \lesssim \varepsilon.
\end{equation}
Intuitively, if \(\SourceNode\) is far from the query point $p_\TargetEdge$ relative to its size, \emph{and} contains tangents that are close together, then the ``lumped'' energy is a good approximation of the total energy between edge $\TargetEdge$  and the edges in \(\SourceNode\).

\paragraph{Differential}

Rather than differentiate our Barnes-Hut approximation of \(\DiscEnergy\), we approximate the differential of the (full) discrete energy directly.  
Starting with the zero vector \(d\smash{\DiscEnergy} = 0 \in \smash{\RR^{3|V|}}\), we perform a BVH traversal for the tangent point \(p_\TargetEdge\) associated with each edge \(\TargetEdge \in E\).  
At each admissible node \(\SourceNode\) and for each endpoint \(\TargetIndex_a\), \(a=1,2\) of $\TargetEdge$ we increment the differential via
\[
   (d\DiscEnergy)_{\TargetIndex_a} 
   +\!=
   \NodeMass_\SourceNode \tfrac{\partial}{\partial \gamma_{\TargetIndex_a}} \left( \Mass_\TargetEdge ( \DiscKernel(\Center_\TargetEdge,\NodeCenter_\SourceNode,\Tangent_\TargetEdge) + \DiscKernel(\NodeCenter_\SourceNode,\Center_\TargetEdge,\NodeTangent_\SourceNode) ) \right).
\]
Here, \(\smash{\DiscKernel}\) is the discrete kernel defined in \eqref{DiscreteKernel}; note that \(\NodeMass_\SourceNode\), \(\NodeCenter_\SourceNode\), and \(\NodeTangent_\SourceNode\) do not depend on \(\gamma_{\TargetIndex_1}\) or \(\gamma_{\TargetIndex_2}\), since \(\TargetEdge\) is not contained in any admissible node \(\SourceNode\).  At any leaf node \(\SourceNode\) we add the corresponding derivatives for all edges \(\SourceEdge \in \SourceNode\).

\subsection{Hierarchical Matrix-Vector Product}
\label{app:HierarchicalMatrixVectorProduct}

\subsubsection{Block Cluster Tree (BCT)}
\label{app:BlockClusterTree}

A BCT partitions a matrix into low-rank \emph{blocks} that approximate the original entries (\figref{BlockClusterTree}).  
It is like a quadtree, except that the matrix ordering is not fixed \apriori.  The basic idea is that the edges in a BVH node \(\Node\) correspond to a subset of BCT rows/columns.  
A block of the BCT is hence specified by a pair of nodes \((\TargetNode,\SourceNode)\) from the BVH.  To construct a BCT, we recursively split the root block \((\RootNode,\RootNode)\), where \(\RootNode\) is the root of the BVH.  
A~block \((\TargetNode,\SourceNode)\) is a leaf if and only if (i) it is \emph{well-separated}, \ie, it provides a good approximation of the local double sum, or (ii) \(\TargetNode\) or \(\SourceNode\) contains just a few edges.  Otherwise, this block is given four children 
\((\TargetNode_1,\SourceNode_1)\), \((\TargetNode_1,\SourceNode_2)\), \((\TargetNode_2,\SourceNode_1)\), \((\TargetNode_2,\SourceNode_2)\), where \(\TargetNode_1,\TargetNode_2\) are the children of \(\TargetNode\) in the BVH (and likewise for \(\SourceNode\)).  
The conditions for being well-separated are similar to \eqref{BarnesHutCondition}:
\begin{equation}
   \label{eq:BCTAdmissibilityTest}
   \frac{\max(r_x^\TargetNode,r_x^\SourceNode)}{|\Center_\TargetNode-\Center_\SourceNode|} 
   \lesssim \varepsilon
   \qquad \text{and} \qquad
   \max( r_T^\TargetNode, r_T^\SourceNode ) 
   \lesssim \varepsilon,
\end{equation}
where \(r_x^\Node\) and \(r_T^\Node\) are the spatial and tangential radii of node \(\Node\).

\subsubsection{Matrix-Vector Product}
\label{app:MatrixVectorProduct}

The BCT is used to accelerate a matrix-vector product \(\varphi = \Ksf\psi\) via the \emph{fast multipole method}.  We adopt the lowest (0th) order version of this method, which is accurate enough for preconditioning.  In particular, for any admissible leaf node \((\TargetNode,\SourceNode)\), the 
midpoints and tangents of edges in \(\TargetNode\) and \(\SourceNode\) are quite coherent relative to the distance between them.  Since the kernel \(k\) is regular, the restriction of \(\Ksf\) to rows \(\TargetEdge \in \TargetNode\) and columns \(\SourceEdge \in \SourceNode\) is hence well-approximated by
\[
   \widehat{\Ksf}_{\TargetNode\SourceNode} := \Index{\Mass}{\TargetNode} \,  k(\NodeTangentPoint_\TargetNode,\NodeTangentPoint_\SourceNode) \, \Index{\Mass}{\SourceNode}^\T,
\]
where \(\Index{\Mass}{\Node} \in \RR^{|\Node|}\) is the vector of edge lengths in $\Node$.  
Using this rank-1 approximation, matrix-vector multiplication amounts to a single dot product (with \(\Index{\Mass}{\SourceNode}\)), followed by a scalar-vector product.

To perform a multiplication, we start with the zero vector \(\varphi = 0 \in \RR^{|E|}\) and iterate over all BCT leaves.  For each admissible leaf \((\TargetNode,\SourceNode)\) (\ie, one which satisfies \eqref{BCTAdmissibilityTest}) we perform an update
\[
   \Index{\varphi}{\TargetNode} \gets \Index{\varphi}{\TargetNode} + \widehat{\Ksf}_{\TargetNode\SourceNode} \, \Index{\psi}{\SourceNode}.
\]
For inadmissible leaves, we simply sum over all edge pairs:
\[
	\varphi_\TargetEdge \gets \varphi_\TargetEdge + \sum_{\SourceEdge \in \SourceNode} \Ksf_{\TargetEdge\SourceEdge} \, \psi_\SourceEdge
\]
for all \(\TargetEdge \in \TargetNode\).  To accelerate evaluation, we percolate these sums up and down the BVH, following a standard fast multipole strategy.

\subsection{Multigrid Solver}
\label{app:MultigridSolver}

We first sketch out a generic multigrid strategy for saddle-point problems on a curve network; the specific solves needed for the tangent-point energy are detailed in \appref{GradientSolveandConstraintProjection}.

\subsubsection{Geometric Multigrid}
\label{app:GeometricMultigrid}

Suppose we want to solve a linear equation \(Ax = b\).  The basic idea of geometric multigrid is to use a coarser mesh to reduce the residual of an equation on the finer mesh.  Consider a simple two-level hierarchy---in particular, let \(\Asf_0 \in \RR^{|V_0| \times |V_0|}\) and \(\Asf_1 \in \RR^{|V_1| \times |V_1|}\) be discretizations of \(A\) on a fine and coarse mesh, \resp{}, and let \(\bsf_0\) be a discretization of the function \(b\) onto the finest mesh.  Also let \(\Jsf_1 \in \RR^{|V_0| \times |V_1|}\) be a so-called \emph{prolongation operator}, which interpolates data from the coarse mesh onto the fine mesh.  Starting with any initial guess \(\xsf_0 \in \RR^{|V_0|}\), we first apply a \emph{smoothing procedure} \(S\) to the system \(\Asf_0 \xsf_0 = \bsf_0\), \ie, a fixed number of iterations of any iterative linear solver to get an improved guess \(\tilde{\xsf}_0 \gets S(\Asf_0,\xsf_0,\bsf_0)\).  We then compute the residual \(\rsf_0 \gets \Asf_0 \tilde{\xsf}_0 - \bsf_0\), and transfer it to the coarse mesh via \(\bsf_1 \gets \Jsf_1^\T \rsf_0\).  On the coarse mesh we solve the system \(\Asf_1 \xsf_1 = \bsf_1\) directly, and transfer the result back to the fine mesh via \(\ysf_0 \gets \Jsf_1 \xsf_1\).  These values are used to update our guess via \(\tilde{\xsf}_0 \gets \tilde{\xsf}_0 + \ysf_0\), and smoothed again.  If the residual is small enough, we stop; otherwise, we repeat another such \emph{V-cycle} until convergence.  More generally, one can apply this two-level strategy to solve the linear system on the coarser level, yielding a multi-level strategy.  The size of the coarsest level is chosen so that a direct solve at this level is more efficient than continuing to apply multigrid.

\paragraph{Initialization.} We get an initial guess \(\xsf_0\) by first coarsening the fine right-hand side \(\bsf_0\) down to the coarsest mesh.  We then perform a direct solve and prolong the solution all the way to the finest mesh, applying smoothing after each refinement.  In practice this strategy works much better than starting with the zero vector.

\paragraph{Implementation Details} In practice we use a standard conjugate gradient smoother, and typically need 6 or fewer V-cycles to achieve a relative residual of order \(10^{-3}\).  Making the residual smaller via further cycles (and a more accurate BCT) yields diminishing returns: we need only a reasonable intermediate descent direction.  Note that although we build a BCT at each level, overall construction cost is only about twice the cost at the finest level.

\subsubsection{Curve Coarsening and Prolongation}
\label{sec:CurveCoarseningandProlongation}

\setlength{\columnsep}{1em}
\setlength{\intextsep}{0em}
\begin{wrapfigure}{r}{68pt}
   \includegraphics{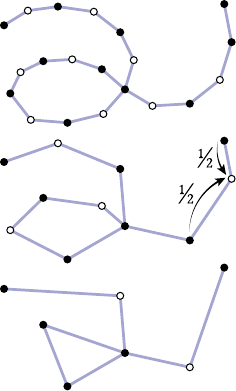}
\end{wrapfigure}
To build a multigrid hierarchy on a general curve network, we apply a simple coarsening scheme.  We mark alternating vertices as ``black'' and ``white'', and mark all endpoints and junctures where two or more curves meet as black.  The next coarsest curve is obtained by removing white vertices, and we stop when we reach a target size or when there are no more white nodes.  The prolongation operator \(\Jsf\) preserves values at black vertices, and at white vertices takes the average of the two neighboring black vertices.  In our experience, using linear interpolation based on edge lengths made no appreciable difference in multigrid performance.  Although coarsening can change the isotopy class of the curve network, it still provides useful preconditioning for the next level of the hierarchy.

\subsubsection{Multigrid for Saddle Point Problems}
\label{app:MultigridforSaddlePointProblems}

Our constraint scheme entails solving \emph{saddle point problems} of the form
\begin{equation}
\label{eq:GenericSaddlePointProblem}
   \left[ \begin{array}{ll}
      \VSobolevMatrix & \ConstraintMatrix^\T \\
      \ConstraintMatrix & 0 \\
   \end{array}
   \right]
   \left[ \begin{array}{c}
      \xsf \\ \uplambda
   \end{array}
   \right]
   =
   \left[ \begin{array}{c}
      \asf \\ 0
   \end{array}
   \right],
\end{equation}
where \(\VSobolevMatrix\) is the inner product (for vector-valued functions) (see \eqref{VectorInnerProduct}), and \(\ConstraintMatrix\) is the constraint matrix (\secref{GradientProjection}); the data \(\asf \in \RR^{3 |V|}\) depends on the problem being solved.  We follow the approach of \citet{Braess:1997:ESS}, who note that for the structurally identical \emph{Stokes' problem} (where \(\VSobolevMatrix\) and \(\ConstraintMatrix\) are replaced by the Laplace and divergence operators, \resp), applying multigrid to the whole matrix does not work well.  Instead, let \(\Psf \in \smash{\RR^{3|V| \times 3|V|}}\) be a projection onto the null space of \(\ConstraintMatrix\), \ie, \(\ConstraintMatrix\Psf = 0\) and \(\smash{\Psf^2} = \Psf\).  Then by construction, any solution \(\ysf\) to the equation
\begin{equation}
   \label{eq:ProjectedSystem}
   \Psf^\T \VSobolevMatrix \Psf \ysf = \Psf^\T \asf
\end{equation}
yields a vector \(\xsf = \Psf\ysf\) within the constraint space \(\Csf\xsf = 0\) that satisfies our original equation.  \eqref{ProjectedSystem} is therefore the system that we actually solve via multigrid.  In particular, we use the projection \(P := \ConstraintMatrix\ConstraintMatrix^\dagger\), where \(\dagger\) denotes the (Moore-Penrose) pseudoinverse
\[
   \ConstraintMatrix^\dagger := (\ConstraintMatrix\ConstraintMatrix^\T)^{-1}\ConstraintMatrix^\T.
\]
Since our constraints are typically sparse, we can factorize the inner term \(\ConstraintMatrix\ConstraintMatrix^\T\) (once per time step) to further accelerate computation.  Note that one must build a constraint matrix \(\ConstraintMatrix_i\) and projection matrix \(\Psf_i\) at each level \(i\) of the multigrid hierarchy.

\subsubsection{Gradient Solve and Constraint Projection}
\label{app:GradientSolveandConstraintProjection}

With these pieces in place, we can apply multigrid to compute the constrained gradient (\eqref{GradientSaddlePointProblem}), and perform constraint projection (\eqref{ConstraintSaddlePointProblem}).

\paragraph{Gradient} To compute the gradient, recall that \(\SobolevMatrix = \LowOrderMatrix + \HighOrderMatrix\).  A~matrix-vector product \(\LowOrderMatrix\usf\) can be expressed as
\begin{equation}
   \label{eq:LowOrderDecomposition}
   \LowOrderMatrix\usf = \Esf^\T(\diag(\Ksf\ones) - \Ksf)\Esf\usf
\end{equation}
where \(\diag(\vsf)\) is a diagonal matrix with entries \(\vsf\), \(\Esf \!\in\! \RR^{|E| \times |V|}\) averages values from vertices to edges (\ie, \((\Esf\usf)_I = \tfrac{1}{2}(u_{i_1}\!\!+\!u_{i_2})\)), and
\begin{equation}
   \label{eq:LowOrderKernel}
   \Ksf_{IJ} = (\Kernel[2][2\sigma+5](\Center_I,\Center_J,\Tangent_I) + \Kernel[2][2\sigma+5](\Center_J,\Center_I,\Tangent_J) )\ell_I \ell_J.
\end{equation}
We use the method from \appref{HierarchicalMatrixVectorProduct} to efficiently perform the products \(\Ksf\ones\) and \(\LowOrderMatrix\usf\), and ordinary sparse matrix multiplication for~\(E\).  The high-order part \(\HighOrderMatrix\) is expressed exactly as in \eqref{LowOrderDecomposition}, except that (i) we replace the averaging operator \(\Esf\) with the difference operator \(\Dsf\), (ii) we define a different kernel matrix \(\Ksf\) by replacing \(\Kernel[2][2\sigma+5]\) with \(\Kernel[0][2\sigma+1]\) in \eqref{LowOrderKernel}, and (iii) just like \(\SobolevMatrix\), \(\Ksf\) acts blockwise on the three components of vector-valued data \(\xsf \in \RR^{3|E|}\) (\ala{} \eqref{VectorInnerProduct}).

\paragraph{Constraint Projection} To use our multigrid solver for constraint projection, we apply a simple transformation to \eqref{ConstraintSaddlePointProblem} that gives it the same form as \eqref{GenericSaddlePointProblem}.  In particular, we solve
\[
   \left[ \begin{array}{ll}
      \VSobolevMatrix & \ConstraintMatrix^\T \\
      \ConstraintMatrix & 0 \\
   \end{array}
   \right]
   \left[ \begin{array}{c}
      \ysf \\ \upmu
   \end{array}
   \right]
   =
   \left[ \begin{array}{c}
      \VSobolevMatrix \zsf \\ 0
   \end{array}
   \right],
\]
where \(\zsf := \ConstraintMatrix^\dagger \bsf\), and \(b\) is the lower block of the right-hand side of \eqref{ConstraintSaddlePointProblem}.  The final result is then given by
\begin{equation}
   \label{eq:SaddleSolutionTransformation}
   \xsf = \zsf - \ysf.
\end{equation}

\end{document}